\documentclass{aa}  
\usepackage{graphicx}
\usepackage{txfonts}
\usepackage{comment}

\usepackage[dvipsnames]{xcolor}
\usepackage{hyperref}
\hypersetup{
     colorlinks = true,
     citecolor = blue,
     }

\usepackage{ulem}
\usepackage{orcidlink}

\begin{document} 

\title{Optical polarization of stellar-fed active and quiescent supermassive black holes}
\titlerunning{Optical polarization of nuclear accretion flares}
\authorrunning{N. Jordana-Mitjans, A. Franckowiak}

\author{N. Jordana-Mitjans  \inst{1}\thanks{\email{jordana@astro.ruhr-uni-bochum.de}}\orcidlink{0000-0002-5467-8277}, A. Franckowiak  \inst{1}\orcidlink{0000-0002-5605-2219}, E. Ramírez-Ruiz\inst{2,3}\orcidlink{0000-0003-2558-3102}, C. G. Mundell\inst{4}\orcidlink{0000-0003-2809-8743}, N. Valtonen-Mattila\inst{1}\orcidlink{0000-0002-1830-098X}, R. Stein\inst{5,6,7}\orcidlink{0000-0003-2434-0387}, \and P. Mil\'an Veres \inst{1}\orcidlink{0000-0002-9553-2987}}

\institute{Ruhr University Bochum, Faculty of Physics and Astronomy, Astronomical Institute (AIRUB), Universitätsstraße 150, 44801 Bochum, Germany.
\and 
Department of Astronomy and Astrophysics, University of California, Santa Cruz, CA 95064, USA
\and 
DARK, Niels Bohr Institute, University of Copenhagen, Jagtvej 128, DK-2200, Copenhagen, Denmark
\and
European Space Agency, European Space Astronomy Centre, E-28692 Villanueva de la Cañada, Madrid, Spain
\and
Department of Astronomy, University of Maryland, College Park, MD 20742, USA
\and
Joint Space-Science Institute, University of Maryland, College Park, MD 20742, USA
\and
Astrophysics Science Division, NASA Goddard Space Flight Center, Mail Code 661, Greenbelt, MD 20771, USA
}


\abstract
    {The advent of wide-field optical surveys with multiwavelength capabilities has been a breakthrough in the detection and characterization of bright, long-lasting optical flares from supermassive black holes (SMBHs), such as tidal disruption events (TDEs), Bowen flares in active galactic nuclei (AGN), and changing-look AGN. Yet, the physical mechanisms powering these diverse transient events remain highly debated.}
    {We aim to provide novel model constraints by utilizing broadband optical polarimetry to study five TDEs. Our goal is to enhance the understanding of the reprocessing material involved in TDEs and to compare our findings with similar flaring activity observed in AGN.}
   {Using the MOPTOP polarimeter on the 2-meter Liverpool Telescope, we conducted a monitoring campaign targeting three optically-discovered TDEs (AT2024bgz, AT2024pvu, and AT2024wsd) and two Bowen flares in AGN (AT2020afhd and AT2019aalc).}
   {The three thermal TDEs showed low intrinsic polarization levels ($\Delta P \approx 0$–$6\%$) with stable polarization angles. The Bowen flares also showed variable polarization degree ($ \Delta P \approx 0$–$8\%$), but significant polarization angle variability: AT2020afhd exhibited a $\Delta \theta = 83 \pm 8 \,\degr$ shift at 150 days post-optical peak, while the AT2019aalc displayed quasi-periodic swings of $\Delta \theta \approx 40 \, \degr$ amplitude starting 190 days after peak brightness. }
    {The TDEs of this study are well described by models invoking rapid disk formation and reprocessed emission from optically thick outflows, whereas the Bowen flares reveal more complex reprocessing geometries, potentially consistent with TDEs occurring in AGN gas-rich environments. We find that moderate polarization  is observed at later times for TDEs with low-Eddington ratios and highly extended photospheres. This implies that, as the accretion level declines, we expect more asymmetric reprocessing layers along a given viewing angle. Since the outflow density and velocity depend sensitively on the inclination angle, we expect TDEs with low-Eddington ratios and highly extended photospheres to exhibit varying levels of polarization. The polarization of  AT2019aalc (Seyfert 1) hints at a clumpy, asymmetric outflow and the presence of a tilted, precessing accretion disk, while the polarization of AT2020afhd (AGN type 2) is consistent with the detection of a scattered light echo.}

   \maketitle

\section{Introduction}
The continuous scan of the sky across wavelengths (e.g., ROSAT; \citealt{1999A&A...349..389V}, eROSITA; \citealt{2021A&A...647A...1P}, Catalina Real-Time Transient Survey; \citealt{2009ApJ...696..870D}, All-Sky Automated Survey for Supernovae; \citealt{2017PASP..129j4502K}, Zwicky Transient Facility; \citealt{2019PASP..131a8002B}, Wide-field Infrared Survey Explorer; \citealt{2010AJ....140.1868W,2011ApJ...731...53M}, Very Large Array Sky Survey; \citealt{2020PASP..132c5001L}) has enabled the detection of nuclear transients, lasting from months to years. The detection of these flares has highlighted the presence of accretion-powered events in a population of previously dormant supermassive black holes (SMBHs), adding to extreme flaring  activity populations that are uncommonly seen  in nearby active galactic nuclei (AGN). Whilst the presence of an accretion disk in AGN can in principle account for such activity (e.g. \citealt{2017ApJ...843L..19M,2019ApJ...874...44Y,2019ApJ...883...94T,2019NatAs...3..242T,2025arXiv250619900D}), flaring activity from quiescent SMBHs is thought to mainly arise mostly from tidal disruption events (TDEs; e.g. \citealt{1988Natur.333..523R,2011ApJ...741...73V,2012ApJ...760..103D,2013ApJ...767...25G,2015JHEAp...7..148K,2016MNRAS.455..859S,2018ApJ...859L..20D,2021ARA&A..59...21G}).

In the dense stellar clusters of galactic nuclei, gravitational interactions among stars are common, placing stars into highly eccentric orbits and leading to possible stellar encounters with the SMBH \citep{2017ARA&A..55...17A}. When the orbit of a (main-sequence) star crosses the tidal radius of an SMBH ($ \lesssim 10^8 M_{\odot}$), the star is ripped apart (e.g. \citealt{1975Natur.254..295H,1978MNRAS.184...87F}), producing streams of bound and unbound debris \citep{1988Natur.333..523R}. In a   full disruption, about half of the stellar-debris mass remains bound to the SMBH and might circularize to form an accretion disk \citep[e.g.,][]{2009ApJ...697L..77R,2014ApJ...783...23G,2015ApJ...804...85S,2015ApJ...809..166G,2016MNRAS.461.3760H}. This accretion disk was initially predicted to emit in the extreme ultraviolet (EUV) and soft X-rays (e.g. \citealt{1988Natur.333..523R}). 

Whilst the first TDE candidates were discovered in the X-ray regime with ROSAT \citep[e.g.,][]{1996A&A...309L..35B,1999A&A...343..775K},  they are now being detected in larger numbers by dedicated wide-field optical surveys, such as the Zwicky Transient Facility (ZTF) ---with advances in the classification algorithms contributing to a discovery rate\footnote{This discovery rate is based on the classification reports submitted to the Transient Name Server (TNS): \url{https://www.wis-tns.org/search}} of $\approx 30$ TDEs year$^{-1}$ \citep{2024ApJ...965L..14S}. However, the thermal properties observed during the first months of these events, coupled with numerous X-ray non-detections, overall X-ray faintness ($L_{\rm X} / L_{\rm opt} < 1$), and delayed X-ray emission, challenge the conventional picture of a newly-formed, compact accretion disk \citep[e.g.,][]{2009ApJ...698.1367G,2011ApJ...741...73V,2019ApJ...878...82V,2020ApJ...889..166J,2022ApJ...925...67L}. One possibility is that, with efficient circularization of the stellar debris, the UV/optical emission results from reprocessing of the inner disk’s X-rays/EUV by optically thick stellar debris, disk outflows \citep{1997ApJ...489..573L,2009MNRAS.400.2070S,2014ApJ...783...23G,2015ApJ...805...83M,2016ApJ...827....3R,2016MNRAS.461..948M,2018ApJ...859L..20D}, or by the wide-angle outflow expected to be produced by stream collisions \citep{2020MNRAS.492..686L,2021MNRAS.504.4885B}. Alternatively, the early UV/optical emission may be powered by shocks at the self-intersection of debris streams during their circularization into an accretion disk \citep{2015ApJ...806..164P,2016ApJ...830..125J,2017MNRAS.464.2816B}.

Polarization has proven a useful tool for constraining the geometry and emission mechanisms of  nuclear accretion flares. TDEs have shown high degrees of polarization from synchrotron radiation emitted in stellar stream shocks ($\Delta P \approx 8-28\%$; \citealt{2023Sci...380..656L}), and low-to-moderate polarization levels in decelerating shocks of relativistic jetted outflows ($\Delta P \approx 0-8\%$; \citealt{2012MNRAS.421.1942W,2020MNRAS.491.1771W}). Thermal emission from TDEs has also exhibited mild polarization levels ($\Delta P \approx 0-6\%$), attributed to electron scattering within the reprocessing layers of the accretion disk, winds, or outflows \citep{2022MNRAS.515..138P, 2022NatAs...6.1193L, 2023A&A...670A.150C, 2024arXiv240304877K}. In these thermal cases, the polarization degree is sensitive to the geometry and optical depth of the reprocessing layer, the clumpiness of the material, the accretion rate, and the viewing angle \citep{2016ApJ...827....3R,2018ApJ...859L..20D,2022ApJ...937L..28T}.

Here, we present the polarization time series of five optically bright TDE candidates. They were selected based on their estimated peak brightness ($r < 18.5$ mag) to enable precise polarization measurements with the Multicolour OPTimised Optical Polarimeter (MOPTOP), and their visibility from the Liverpool Telescope for a minimum duration of one month. The selected events are AT2024bgz (with redshift $z=0.0585$; \citealt{2024TNSCR.401....1G}), 
AT2024pvu ($z=0.0488$; \citealt{2024TNSAN.221....1S}), 
AT2024wsd ($z=0.063$; \citealt{2024TNSAN.302....1S}),
AT2020afhd ($z=0.027$; \citealt{2024TNSAN..53....1A}), 
and the second flaring episode of AT2019aalc ($z=0.0356$; \citealt{2024arXiv240817419V}). The AT2019aalc is hosted by a Seyfert type-1 AGN galaxy \citep{2014ApJ...788...45T}, AT2020afhd by an AGN type 2 \citep{2022ApJS..258...29C}, and they have both been classified as Bowen fluorescence flares \citep{2024arXiv240817419V,2024TNSAN..53....1A}. Such flares have long-lasting TDE-like UV/optical emission, with emission lines typical of unobscured AGN. However, they also show broad Bowen fluorescence emission lines (O III $\lambda 3133$, N III $\lambda 4640$, and He II $\lambda 4686$) that are very rare in AGN \citep{1990ApJ...362...74S}. These emission lines are thought to arise from fast, high-density gas that is rich in metals near the accreting SMBH; they have been proposed to be either due to bright-UV transient emission ionizing the broad-line region or a newly-launched outflow \citep{2019NatAs...3..242T}. Comparing our polarimetric results with previous measurements, we aim to constrain the emitting mechanisms of these five nuclear accretion flares. Additionally, we seek to determine whether Bowen flares are stellar-fed SMBHs (i.e., TDEs) or enhanced accretion in AGN (e.g. an unknown form of AGN instability).

This paper is organized as follows. In Section~\ref{sec:observations}, we describe the polarimetric observations and data reduction procedures (see Appendix \ref{appendix_data_red} for more details). The results of our polarization measurements are presented in Section~\ref{sec:results}. In Sections~\ref{sec:TDEs} and \ref{sec:BowenFlares}, we discuss the implications of the polarization time series for the three thermal TDEs and two Bowen flares, respectively. Finally, we summarize our main conclusions in Section~\ref{sec:conclusions}. Throughout this work, we adopt a flat $\Lambda$CDM cosmology with $\Omega_{\rm m} = 0.32$, $\Omega_{\Lambda} = 0.68$, and $H_0 = 67$ km s$^{-1}$ Mpc$^{-1}$  \citep{2020A&A...641A...6P}.

\section{Observations and data reduction} \label{sec:observations}
The MOPTOP is the fourth generation polarimeter at the 2-m fully autonomous Liverpool Telescope \citep{2004AN....325..519S,2020MNRAS.494.4676S}, located at Roque de Los Muchachos Observatory (La Palma, Spain). In the MOPTOP instrumental configuration, the light beam goes through a continuously rotating half-wave plate and a wire-grid polarizing beamsplitter, which splits the light beam into two orthogonal polarizing states (e.g. 0$\degr$ and 90$\degr$) that are simultaneously recorded by two identical CMOS sensors. Each time that the half-wave plate is rotated by 22.5 degrees, the light beam's polarization angle changes by 45 degrees, and the modulated intensity is recorded by the CMOS. A complete rotation of the half-wave plate has 16 steps, which yields a total of 16 images per CMOS. 

The MOPTOP allows filtered observations across optical and near-infrared bands. To maximize the signal-to-noise ratio of the TDEs, we used the broad-band MOPTOP filter (MOP-L) covering $400-700\,$nm. Given the brightness of AT2019aalc and AT2020afhd, we also used the MOP-I filter to cover the near-infrared ($695-950\,$nm) for these sources. The MOPTOP half-wave plate has two rotor speed options: an exposure time of 0.4 sec or 4 sec per frame. To minimize the readout noise, we used an exposure time of 4 sec per image. In this way, the polarization degree and angle of the target were measured every 8 seconds. Observations of the five TDE candidates were scheduled through the Liverpool Telescope user interface phase2UI\footnote{\url{https://telescope.livjm.ac.uk/PropInst/Phase2/}}. They were automatically dispatched by the Liverpool Telescope autonomous scheduler under good observing conditions; that is, we placed observing constraints on the seeing ($<2 \arcsec$) as well as the Moon's phase and distance (see Table \ref{table:log}). The targets were observed for $N_{\rm rot}$ full rotations of the half-wave plate, with $N_{\rm rot}$ adjusted depending on the faintness of the TDE candidate (see Table~\ref{table:log}). The data was automatically processed using the Liverpool Telescope reduction pipeline, which corrects for bias, darks, and flats, and then retrieved from the Liverpool Telescope archive. 

With the MOPTOP, polarization is measured by extracting the counts of a source at each image. Then, from the relative intensities across the different positions of the half-wave plate, one can derive the Stokes parameters ($q,u$). Instead of measuring the polarization directly from the default 4-sec frames, we improved the signal-to-noise ratio of the target by stacking frames. We aligned and then combined the frames to improve the target's point-spread function profile (see Appendix \ref{max_SNR}). We used the Astropy Photutils package to compute the source counts with aperture photometry \citep{2016zndo....164986B}. From source counts, we followed the procedure of \cite{2020MNRAS.494.4676S} to normalize the two detector arrays and we derived the Stokes $q$ and $u$. We corrected the Stokes parameters for instrumental polarization and depolarization using standard stars (see Appendix \ref{pol_inst_all}). The confidence levels of the Stokes parameters ($q$, $u$) and polarization degree ($P$) and angle ($\theta$) were derived from a Monte Carlo error propagation, starting from $10^{6}$ simulated count values per half-wave plate position and CMOS (see e.g. \citealt{2021MNRAS.505.2662J}). The polarization degree $P_{\rm obs}= 100 \sqrt{q^2 + u^2}$ is biased to larger values in the low signal-to-noise ratio regime \citep{1985A&A...142..100S}. Consequently, the polarization degree was corrected using the maximum probability estimator $P= (P_{\rm obs} - \sigma_{ P} ^2/P_{\rm obs})$ from \cite{1997ApJ...476L..27W}, where $\sigma_{P}= 100 \sqrt{(q^2 \sigma_{q}^2 + u^2 \sigma_{u}^2)/(q^2 + u^2)}$ is the $1\sigma$ uncertainty in $P_{\rm obs}$.

\section{Results} \label{sec:results}

After correcting for instrumental effects (see Appendix \ref{pol_inst_all}), here we detail the subsequent corrections applied to recover the intrinsic polarization of the five transients; specifically how we removed the Galactic ISM and host galaxy contributions.

\subsection{ISM dust-induced polarization}
Dust grains align perpendicularly to the local magnetic field and absorb light more efficiently along their longer axes through dichroic extinction (e.g. \citealt{1975ApJ...196..261S,2004ASPC..309...65W}). As a result, when light passes through the ISM, the transmitted light becomes polarized, and the polarization angle traces the
direction of the local magnetic field. This effect is wavelength-dependent, determined by both the dust grain distribution and the efficiency of the alignment. For extragalactic targets, the Galactic dust can induce significant levels of polarization in lines of sight affected by extinction.

To correct for that, we measured the polarization of bright stars (< 14.4 mag) within the MOPTOP field of view ($7 \arcmin \times 7\arcmin$) as a proxy for the polarization contribution from Galactic dust birefringence in each target line of sight (see Table~\ref{table:prop_env}). Following,  we checked whether our Galactic foreground estimates are compatible with those predicted by the Serkowski law \citep{1975ApJ...196..261S,1992ApJ...386..562W}, in which the polarization degree follows $P_{\rm serk} = P_{\rm 0, \, serk} \exp \big[-K \ln^2  \big( \lambda_{\rm max} / \lambda \big)\big]$. Here, the $\lambda_{\rm max}({\rm \mu m})=R_{V}/5.5$ is the wavelength at which the polarization is maximum, $R_{V}=3.1$ is the Galactic ratio of visual extinction to reddening, and $K = (0.01 \pm 0.05) + (1.66 \pm 0.09)\lambda_{\rm max}$ is a constant. Given the low column densities of Galactic hydrogen in the targets' lines of sight (N$_{\rm H} < 5 \times 10^{20}\,{\rm cm}^{-2}$), the upper limit on the polarization degree is $P_{\rm 0, \, serk}  \lesssim 13 \% \, E(B-V)_{\rm MW}$ \citep{2020A&A...641A..12P}, where $E(B-V)_{\rm MW}$ is the Galactic reddening \citep{2013MNRAS.431..394W}. As shown in Table~\ref{table:prop_env}, both the stellar measurements and the reddening-based upper limits are in good agreement. We then subtracted the Galactic dust-induced polarization from the measured polarization as $(q,u)_{\rm TDE+host} = (q,u) - (q,u)_{\rm MW}$. The corrected polarization measurements are presented in Table~\ref{table:pol_phot_resultats}.

\begin{figure*}
\begin{center}
\includegraphics[width=0.45\textwidth]{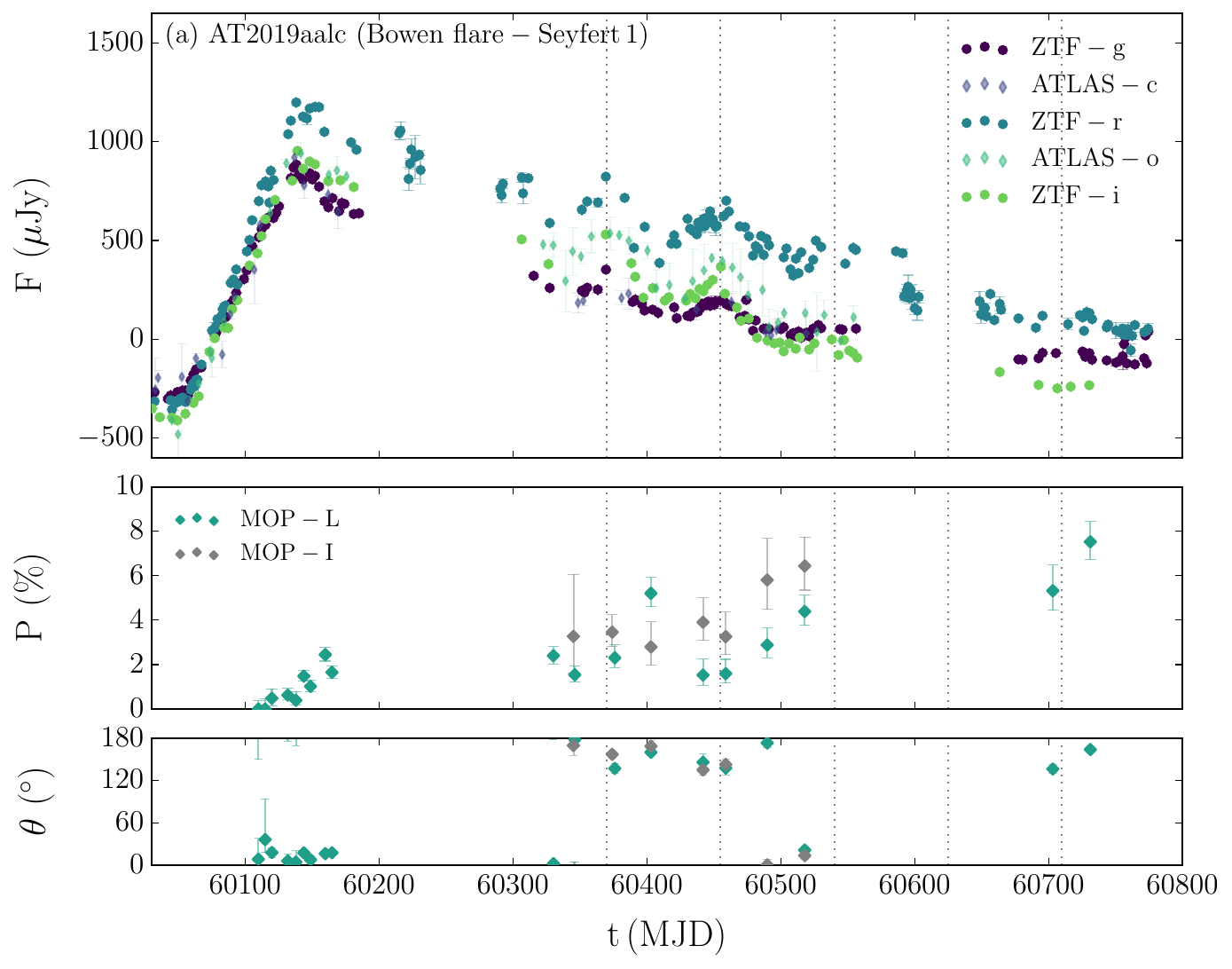}
\includegraphics[width=0.45\textwidth]{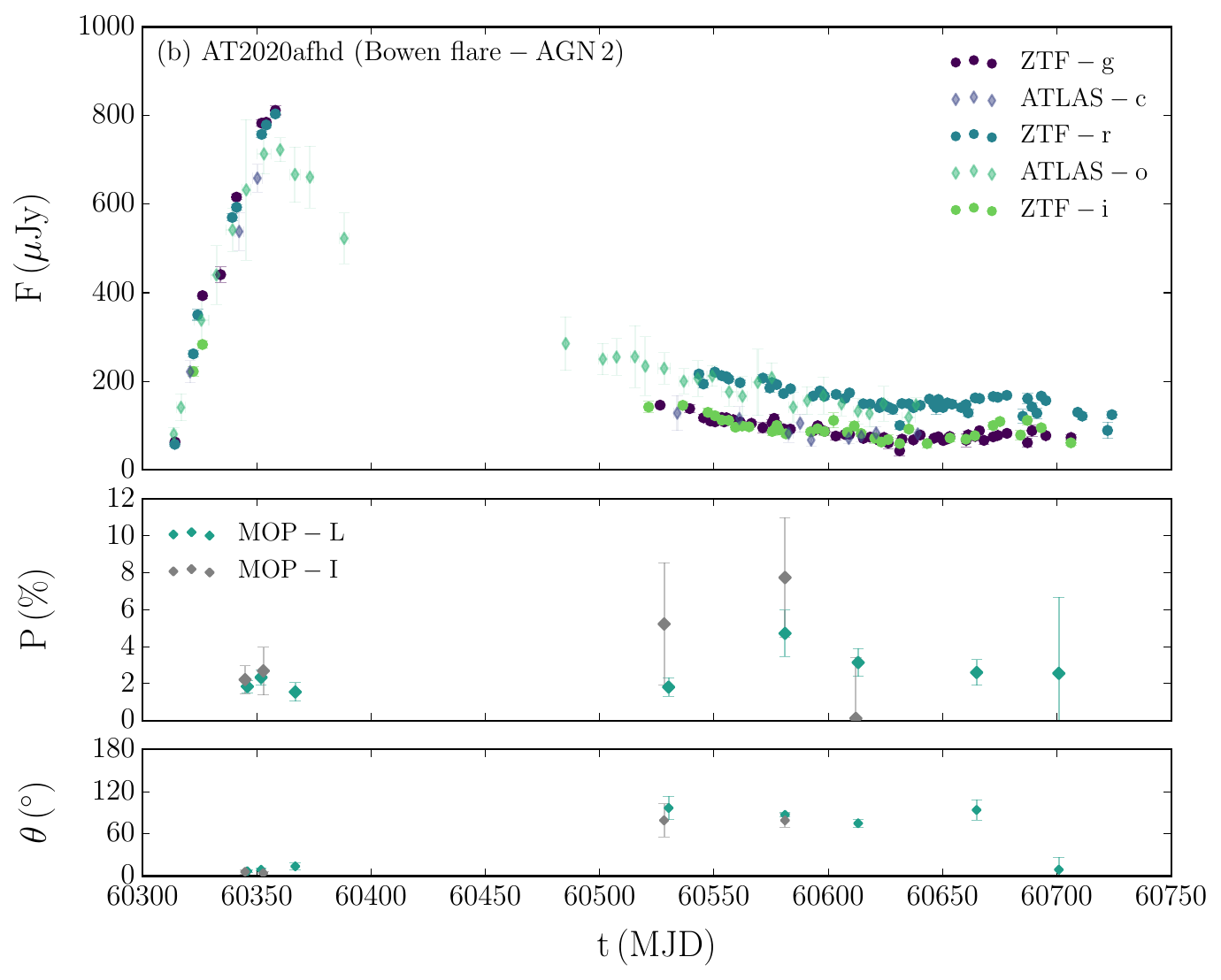}
\includegraphics[width=0.3\textwidth]{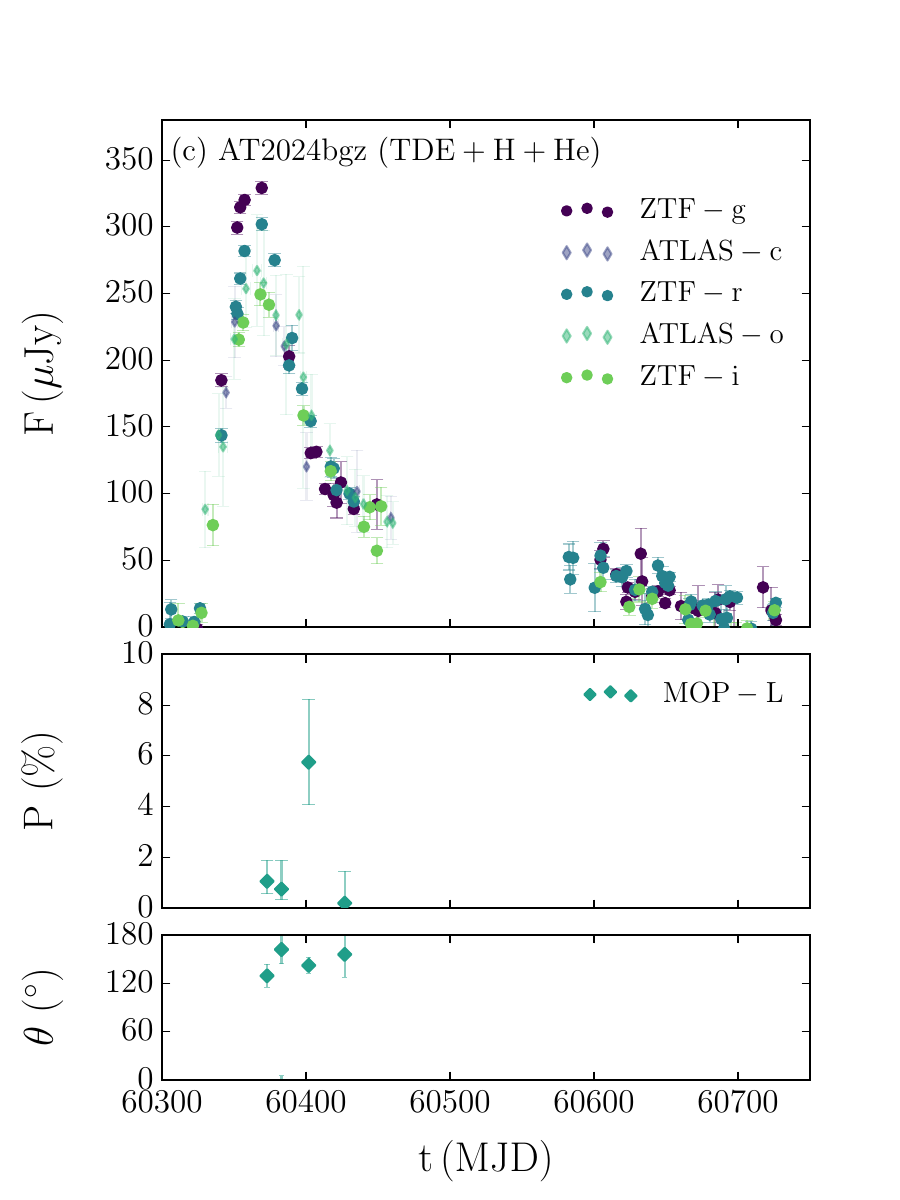}
\includegraphics[width=0.3\textwidth]{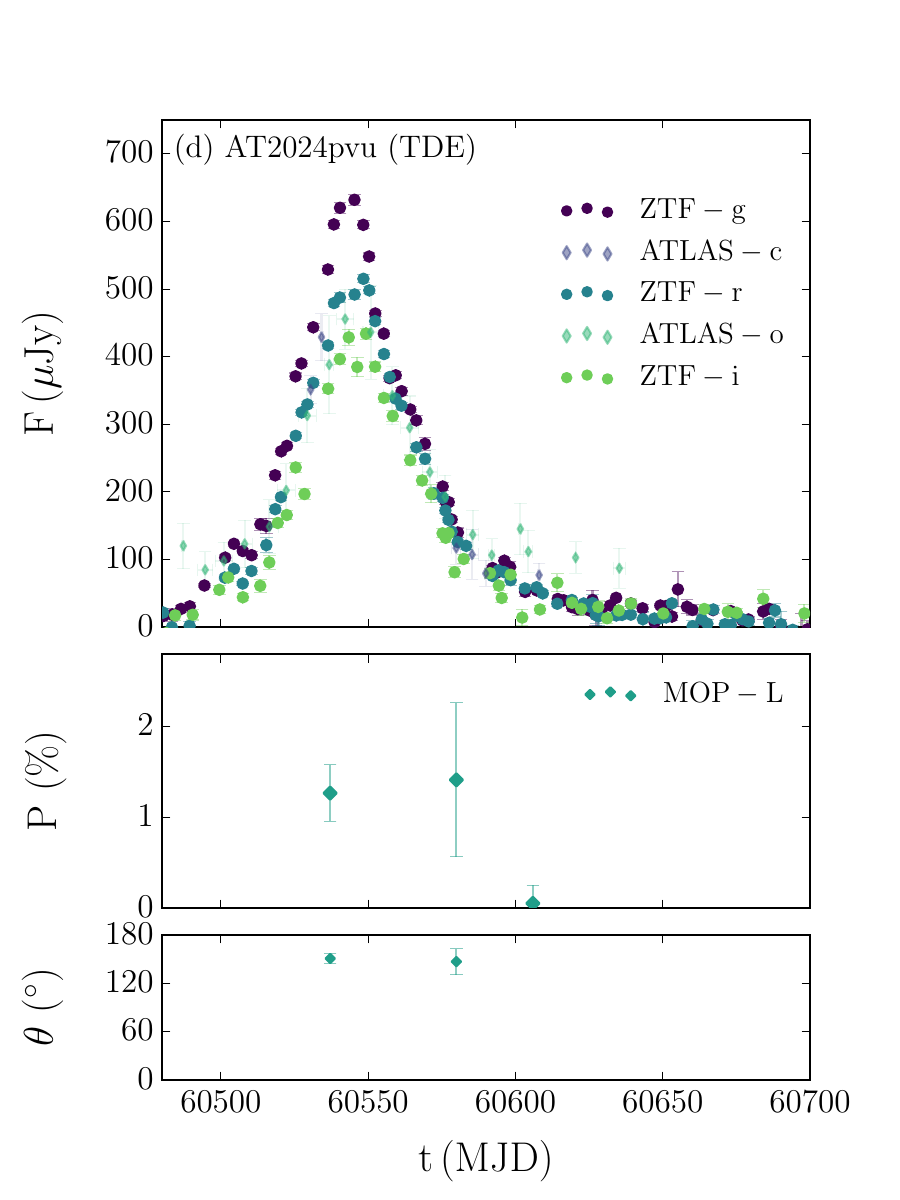}
\includegraphics[width=0.3\textwidth]{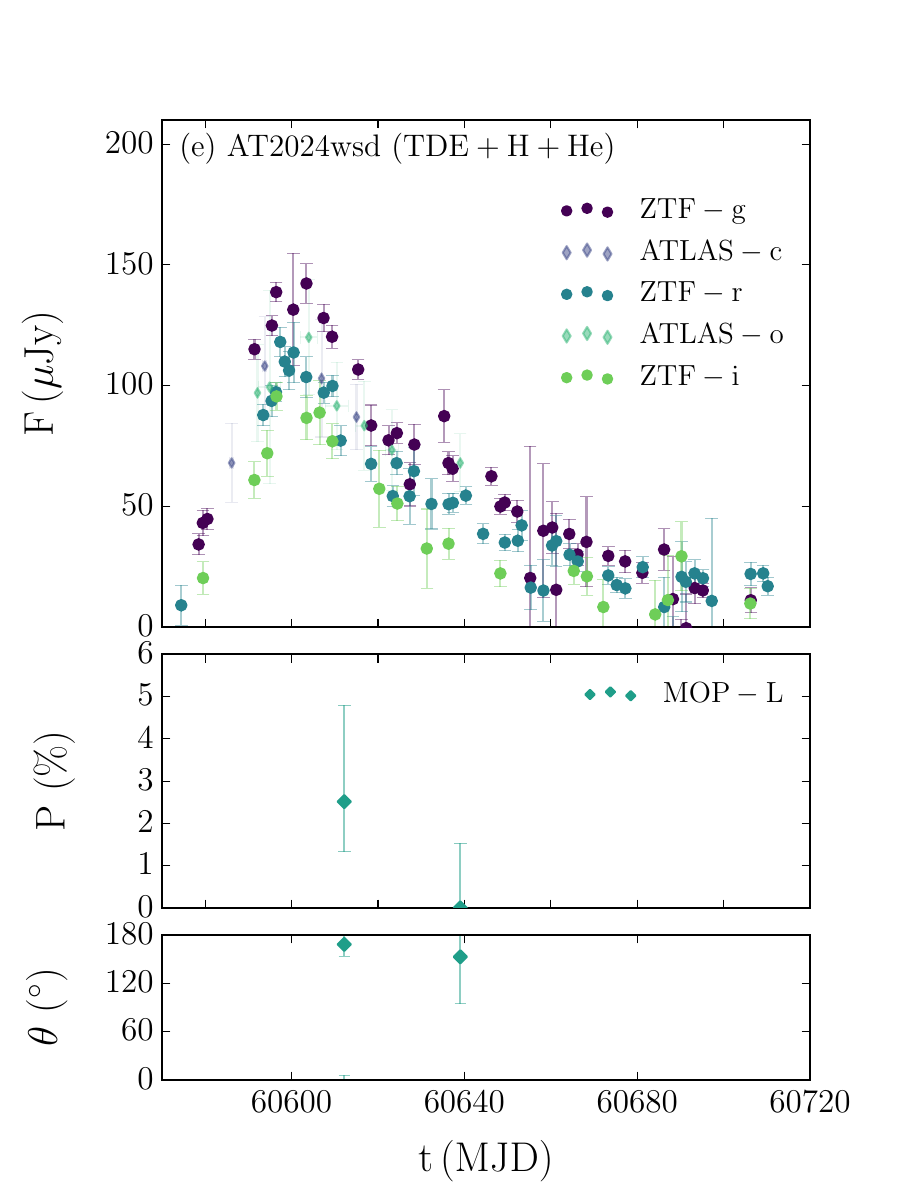}
\end{center}
\caption{ZTF and ATLAS \citep{2018PASP..130f4505T,2021TNSAN...7....1S} optical light curves of (a) AT2019aalc, (b) AT2020afhd, (c) AT2024bgz, (d) AT2024pvu, and (e) AT2024wsd, for filters ZTF $g$ ($420-540 \,$nm), $r$ ($570-720 \,$nm), and $i$ ($720-880 \,$nm), and ATLAS $c$ ($420-650 \,$nm) and $o$ ($560-820 \,$nm). The ZTF data have been binned into fixed time windows of 7 days in the observer rest frame. In the bottom panels, we show the MOPTOP polarization in the MOP-L ($400-700 \,$nm) and MOP-I filters ($700-1000 \,$nm) corrected for Galactic ISM and the host galaxy. Note that the polarization angle $\theta = 0 \degr$ is equivalent to $\theta = 180 \degr$. For AT2019aalc, the vertical dotted black lines correspond to the flux maxima of the quasi-periodic oscillations (QPOs); see details in Section \ref{sec:QPOs}.}
\label{fig:Pol_LC}
\end{figure*}

\subsection{Host galaxy depolarization} \label{sec:depo}
After correcting for Galactic ISM, the polarization of the five nuclear transients still includes a significant contribution of background photons from the host galaxy (i.e. $45-65 \%$ at peak time), which depolarizes the overall measurement. Therefore, we further corrected the polarization by flux-weighting the host and transient polarization as $(q,u)_{\rm TDE} = ((q,u)_{\rm TDE+host} - (1- C_{\rm F}) (q,u)_{\rm host})/C_{\rm F}$, where $C_{\rm F} = F_{\rm TDE}/(F_{\rm TDE} + F_{\rm host})$ is the ratio between the TDE flux and the total measured flux, which includes the galaxy host. Note that we extracted the MOPTOP photometry of the transients using a fixed full width at half maximum (FWHM) for the photometric aperture, including the host galaxy's flux; hence $F_{\rm host}$ is constant across observations. To measure the coefficient $C_{\rm F}$ for each polarimetric data point, we used pre-flare photometric data from the ZTF survey \citep{2019PASP..131a8002B}. To cover the spectral range of the MOPTOP filters, we used a weighted average of ZTF $g$ and $r$ filters for MOP-L observations and the ZTF $i$ filter for MOP-I observations.

The polarization degree of AT2024bgz and AT2024wsd were consistent with zero in the final observing epoch; therefore, we computed the intrinsic polarization of AT2024bgz and AT2024wsd solely assuming starlight dilution in their host galaxies ($P_{\rm host}\approx 0 \%$), such that $P_{\rm TDE} = P_{\rm TDE+host}/C_{\rm F}$ (e.g., \citealt{2022NatAs...6.1193L,2023Sci...380..656L}). However, we could not assume an unpolarized host for AT2024pvu, since in the latest polarimetric epoch (when the TDE contributed only $6\%$ of the total flux) the measured polarization was still non-zero, indicating that the host galaxy itself is polarized ($P_{\rm host} \approx 0.6\%$). Note that non-negligible polarization degrees in quiescent galaxies has been attributed to dichroic absorption of aligned dust grains \citep{1991MNRAS.252..288B}. Therefore, we subtracted the host baseline as a flux-weighted vector component from the total polarization of AT2024pvu.

Initially, AT2019aalc showed low polarization degree ($P < 0.2 \%$ at $2\sigma$), which is consistent with a near-polar viewing angle of a Seyfert type-1 galaxy \citep{2012A&A...548A.121M,2014MNRAS.441..551M, 2014ApJ...788...45T}. Consequently, we estimated the flare intrinsic polarization by assuming that the host and AGN emission are unpolarized ($P_{\rm host} \approx 0 \%$; \citealt{2018A&A...615A.171M}). Given that the AGN continuum decreased $\approx 0.2 \,$mag after the first flare, we adopted $59500-60050$ MJD as the average reference epoch to calculate the baseline flux. In contrast, AT2020afhd is hosted by an AGN type 2 and meets the classification criteria for both a Seyfert 2 galaxy and a type-2 low-ionization nuclear emission-line region (LINER; \citealt{2022ApJS..258...29C}). Initially, AT2020afhd exhibited a higher polarization degree than AT2019aalc ($P \approx 1.5\%$), in line with what is expected by a type‑2 AGN viewed by a mid-latitude observer \citep{2014MNRAS.441..551M}. To estimate the host and AGN polarization correction, we used the epoch when the flare accounted for only $8\%$ of the total MOP-I band flux as a reference, adopting a host polarization of $P_{\rm host} \approx 1.0 \%$.

\subsection{Intrinsic polarization}
In Figures \ref{fig:Pol_LC} and Table \ref{table:pol_phot_resultats}, we present the results corresponding to the intrinsic polarization of the five nuclear transients. The AT2024bgz, AT2024wsd, and AT2024pvu were intrinsically polarized at $\Delta P \approx 0-6\%$ level, with higher polarization levels detected within the first 40 days after optical-peak brightness; after that, the polarization degree is consistent with zero. The polarization angle is constant within $1\sigma$, consistent with an axisymmetric geometry. AT2019aalc and AT2020afhd showed variable polarization levels in the range of $\Delta P \approx 0-8\%$. The polarization degree of AT2019aalc initially increased from $P \approx 0 \% $ at 30 days before the peak to $P \approx 2 \%$ at 25 days after the optical peak, with an average polarization angle $\theta \approx 12 \,\degr$. From 190 days post-peak, the polarization degree and angle in both MOPTOP filters showed variability modulated by the optical rebrightenings, with amplitude $\Delta P \approx 5\%$ and $\Delta \theta \approx 40 \, \degr $; see Sections \ref{sec:QPOs} and \ref{sec:AT2019aalc_late}. The time-series data of AT2020afhd exhibited variable polarization degree at $\Delta P \approx 0-8 \%$ level. Additionally, a striking rotation of the polarization angle by $\Delta \theta = 83 \pm 8 \,\degr$ was observed at $\approx 150 \,$days after optical peak brightness.

\section{Classical TDEs: AT2024bgz, AT2024pvu, and AT2024wsd} \label{sec:TDEs}

Here, we discuss the properties  of optical polarization of AT2024bgz, AT2024pvu, and AT2024wsd in terms of thermal TDEs from quiescent SMBHs, and we compare them with the time-series polarization data of other TDEs. 

\subsection{Spectral properties of thermal TDEs}

\begin{figure}
\begin{center}
\includegraphics[width=0.5\textwidth]{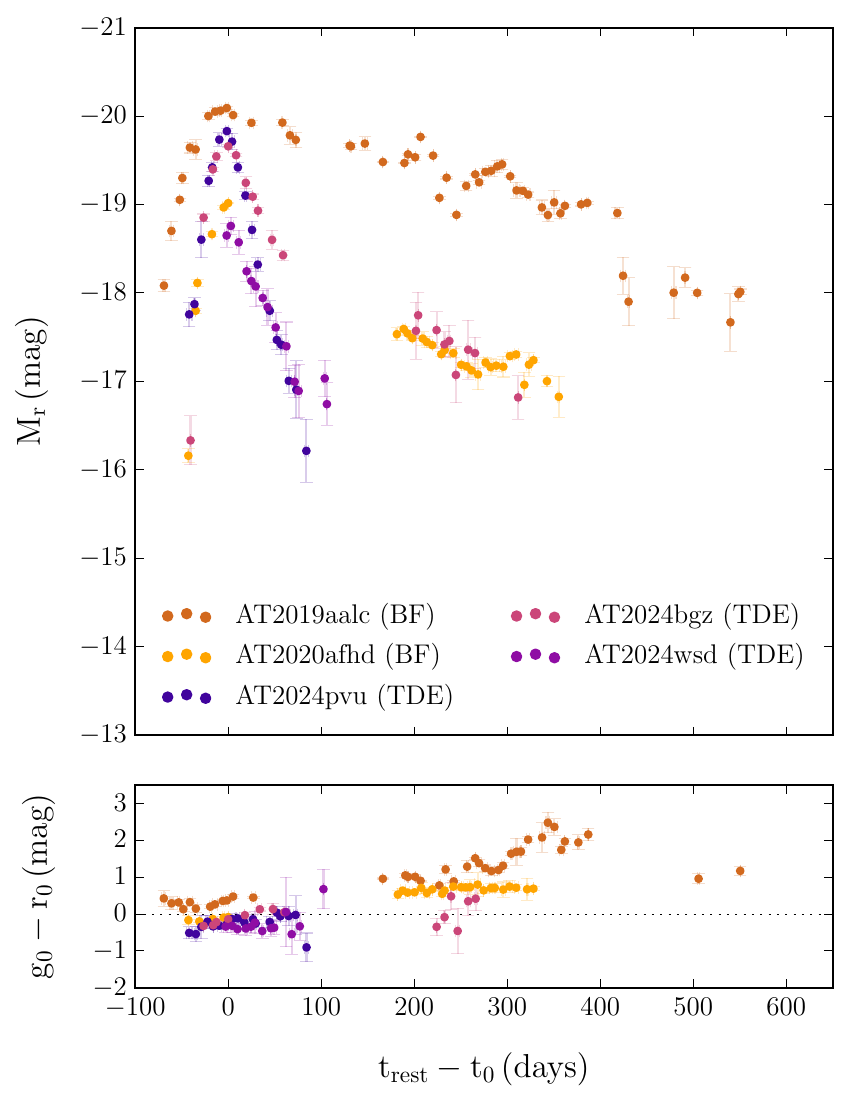}
\end{center}
\caption{ZTF $r$-band light curves of the TDE candidates analysed in this work, binned to 7 days in the rest frame and time-shifted to optical peak brightness ($t_0$). The absolute magnitudes have been calculated as $M_{\rm r} = m_{\rm r} - 5 \log_{10}(d/10) - K(z) - A_{\rm r}$, where $m_{\rm r}$ is the apparent magnitude, $d$ is the luminosity distance in parsecs, $A_{\rm r}$ accounts for Galactic extinction, and we have assumed a uniform $K$-correction, $K = 2.5 \log_{10} (1 + z)$. The bottom panel shows the colour evolution corrected for Galactic extinction and host galaxy contribution.}
\label{fig:LC_abs}
\end{figure}

Thermal TDEs show a fast increase of UV/optical flux, high luminosities and temperatures \citep{2021ApJ...906..101M}, and large blackbody radii, which distinguishes them from typical AGN variability and other transients \citep{2011ApJ...741...73V,2014ApJ...793...38A,2018ApJS..238...15H,2018ApJ...852...37A,2019ApJ...872..151M,2021SSRv..217...63V}. Using ZTF forced photometry\footnote{\url{https://irsa.ipac.caltech.edu/data/ZTF/docs/ztf_forced_photometry.pdf}} (see Figure \ref{fig:LC_abs}), we found that AT2024bgz, AT2024pvu, and
AT2024wsd are consistent with TDE typical luminosities (M$_{\rm r} \approx -19.5$; \citealt{2023ApJ...942....9H}), blue continuum ($g_{\rm 0} -r_{\rm 0}  \approx -0.2$), little colour evolution during months ($\delta (g_{\rm 0} -r_{\rm 0})  <0.006 \, $mag day$^{-1}$; \citealt{2023ApJ...955L...6Y}), and typical spectral classes. That is, the optical spectra of most thermal TDEs show very broad emission lines that become narrower and fainter with time (${\rm FWHM} \, \approx 10^4 \,$km s$^{-1}$; \citealt{2015JHEAp...7..148K,2021ARA&A..59...21G}). Most TDEs show a very pronounced broad helium emission line (He II $\lambda$4686), a feature that sets them apart from other transients (e.g. \citealt{2012Natur.485..217G}). TDEs with only this characteristic spectral feature visible are usually classified into the TDE-He spectral class (e.g. \citealt{2020SSRv..216..124V}). Other TDE spectral classes include TDEs with only H$\alpha$ and H$\beta$ hydrogen emission lines (i.e. the TDE-H class), those with both the helium and hydrogen emission lines (TDE-H+He), and TDEs with featureless spectra. In this context, the spectra of AT2024wsd\footnote{\url{https://www.wis-tns.org/object/2024wsd}} and AT2024bgz\footnote{\url{https://www.wis-tns.org/object/2024bgz}} showed broad H$\alpha$ and He II emission, with AT2024wsd also displaying H$\beta$ ---features typical of the TDE-H+He spectral class \citep{2024TNSAN..47....1G,2024TNSAN.302....1S}. In contrast, AT2024pvu exhibited TDE-like continuum without detectable broad emission lines\footnote{\url{https://www.wis-tns.org/object/2024pvu}} \citep{2024TNSAN.221....1S}.

TDEs-H+He have lower rest-frame luminosities, but higher intrinsic disruption rates compared to the rest \citep{2021ApJ...908....4V,2022MNRAS.515.5604N}. To date, most thermal TDEs with available polarimetric observations belong to this H+He class \citep{2021ApJ...908....4V,2022MNRAS.515.5604N,2023ApJ...942....9H}, including AT2018dyb, AT2019dsg, AT2019azh, AT2019qiz, and AT2020mot \citep{2020ApJ...892L...1L,2022NatAs...6.1193L,2022MNRAS.515..138P,2023Sci...380..656L}. This class has shown clear signatures of Bowen fluorescence lines (i.e. O III and N III; \citealt{2019ApJ...873...92B,2021ApJ...908....4V,2023ApJ...942....9H}) indicating the presence of high-density, high-velocity gas outflows near the SMBH. The relatively small photospheric radii inferred for these events have been proposed to explain the high densities required for efficient Bowen fluorescence and suggested that TDEs-H+He could be due to the disruption of low-mass stars \citep{2021ApJ...908....4V}. Alternatively, these high densities could be achieved by the rapid formation of an accretion disk, driven by stellar debris streams intersecting close to the SMBH in deep encounters \citep{2022MNRAS.515.5604N}. Using MOSFiT, we modelled the ZTF light curves of AT2024bgz and AT2024wsd \citep{2018ApJS..236....6G,2019ApJ...872..151M}; we found that both TDEs are well explained by small photospheric radii and deep encounters of low-mass stars ($\approx 0.3 M_{\odot}$) with $\approx 3 \times 10^6 M_{\odot}$ SMBHs (see Table \ref{tab:mosfit}). We note that, so far, only AT2020mot has shown polarimetric evidence of shock-powered emission \citep{2023Sci...380..656L}. Our MOSFiT modelling of AT2020mot returned a photospheric radius comparable to other TDEs-H+He and suggested a shallow encounter of a Sun-mass star with a $\approx 10^7 M_{\odot}$ SMBH (see Table \ref{tab:mosfit}). Finally, our modelling of the featureless TDE AT2024pvu indicates a photosphere at least an order of magnitude larger than that of TDEs-H-He at explosion time, consistent with a Sun-mass star tidally disrupted by a $\approx 2 \times 10^7 M_{\odot}$ SMBH.

\subsection{TDE optical emission from reprocessing vs. shocks}

\begin{figure}
\begin{center}
\includegraphics[width=0.5\textwidth]{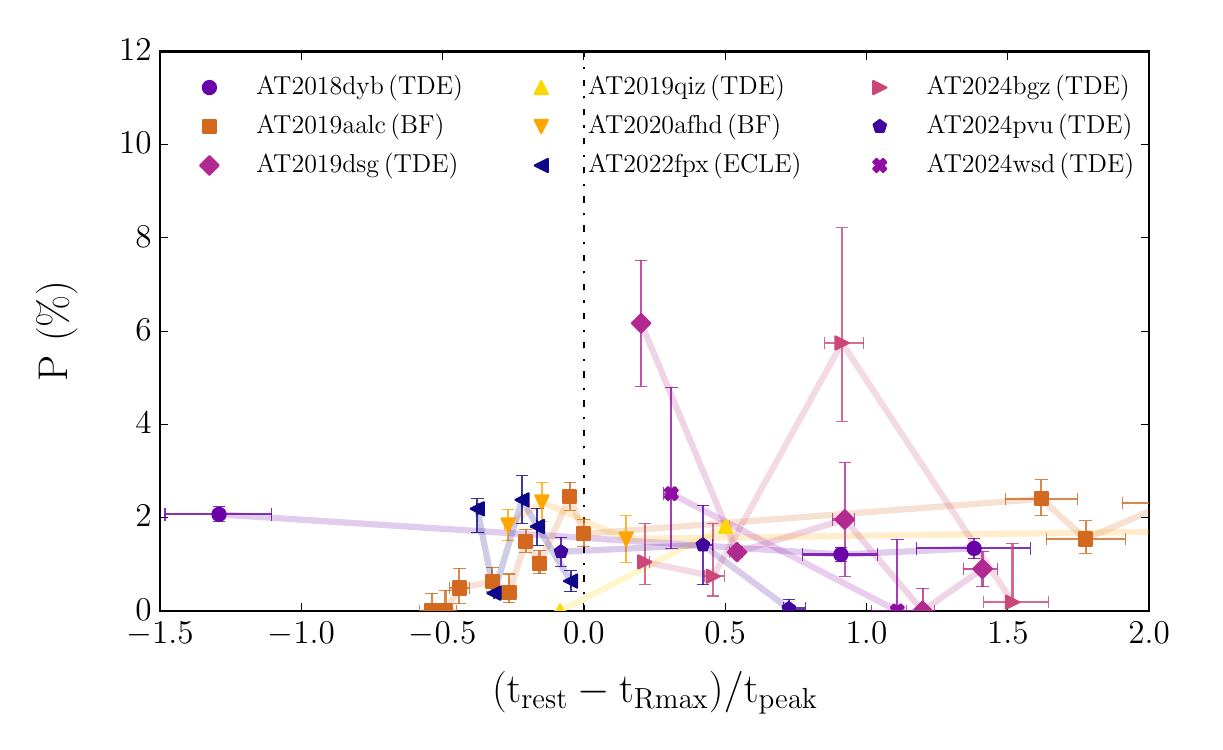}
\end{center}
\caption{Rest-frame polarization measurements of thermal TDEs \citep{2022MNRAS.515..138P,2022NatAs...6.1193L}, Bowen flares (BFs), and the extreme coronal line emitter AT2022fpx (ECLE; \citealt{2024arXiv240304877K}). Time is scaled by each event’s rise time to peak bolometric luminosity to remove the primary timing dependencies on SMBH and stellar mass (i.e., $t_{\rm peak} \propto M_{\rm BH}^{1/2} M_{\star}^{-1} R_{\star}^{3/2}$). Here, the $t_{\rm Rmax}$ is the epoch at which the MOSFiT-derived blackbody radius (and, to a good approximation, the bolometric luminosity) reaches its maximum. Polarization measurements at $\Delta t/t_{\rm peak} < 0$ correspond to the optically-thick expansion regime of the electron-scattering photosphere, where polarization reflects the asymmetry of the outermost layers; the dotted line at $\Delta t/t_{\rm peak} = 0$ marks the photospheric turnover at $\Delta  t/t_{\rm peak} > 0$, where the photosphere recedes as the ejecta becomes increasingly more optically thin. This allows asymmetries in the inner ejecta to become more apparent in their polarization imprint. Note that AT2019qiz  has been corrected by an estimation of host depolarization using ZTF data and assuming $P_{\rm host} \approx 0 \%$.}
\label{fig:Pol_restframe_TDEs}
\end{figure}

The polarization time series of AT2024bgz, AT2024pvu, and AT2024wsd closely resemble those of AT2019dsg and AT2018dyb \citep{2020ApJ...892L...1L,2022NatAs...6.1193L}. That is, constant polarization angle and mild levels of polarization degree, with an eventual decrease to negligible values at late times (see Figure \ref{fig:Pol_restframe_TDEs}). This behaviour is consistent with emission dominated by reprocessing in an optically-thick medium \citep{2016ApJ...827....3R}, rather than shocks in colliding stellar streams, which have been proposed to explain the higher polarization levels and variability in the polarization angle observed in AT2020mot ($\Delta P \approx 8-28\%$, $\Delta \theta \approx 30 \degr$; \citealt{2023Sci...380..656L}).

In a reprocessing scenario, \cite{2022NatAs...6.1193L} proposed that the polarization signatures of AT2019dsg\footnote{Initially, the polarization of AT2019dsg was interpreted as potentially arising from a relativistic TDE outflow \citep{2020ApJ...892L...1L}; however, subsequent multiwavelength observations did not support this scenario \citep{2021ApJ...919..127C}.} and AT2018dyb were due to the rapid circularization of an initially non-axisymmetric accretion disk. Using the super-Eddington accretion model of \citet{2018ApJ...859L..20D}, where an optically, geometrically-thick disk and outflows reprocess the X-ray emission from the inner accretion disk, polarization can reach up to $P \approx 6\%$ for edge-on views of high-mass extended disks. In this model, the polarization degree decline with time is driven by a decreasing fallback rate, leading to a drop in the disk mass and, consequently, a reduction in the electron scattering opacity. Alternatively, \citet{2023A&A...670A.150C} proposed that the diverse behaviour of polarization in TDEs (including the increasing polarization degree of AT2019qiz; \citealt{2022MNRAS.515..138P}) could arise from the reprocessing of a wide-angle outflow generated by the collision at the self-intersection of the returning debris streams \citep{2020MNRAS.492..686L}. Owing to its higher degree of asymmetry, this lower mass and more confined outflow can produce polarization degrees up to $P \approx 11 \%$ for intermediate viewing angles. The early high-density outflow would result in multiple electron scatterings, leading to almost unpolarized emission as directional information is likely to be lost. As the optical depth decreases, non-zero polarization emerges, influenced primarily by the mass outflow rate, geometry, and viewing angle. Eventually, as the ejecta becomes very optically thin, photons escape without further scattering, causing the polarization to disappear. 

\begin{figure*}
\begin{center}
\includegraphics[width=0.45\textwidth]{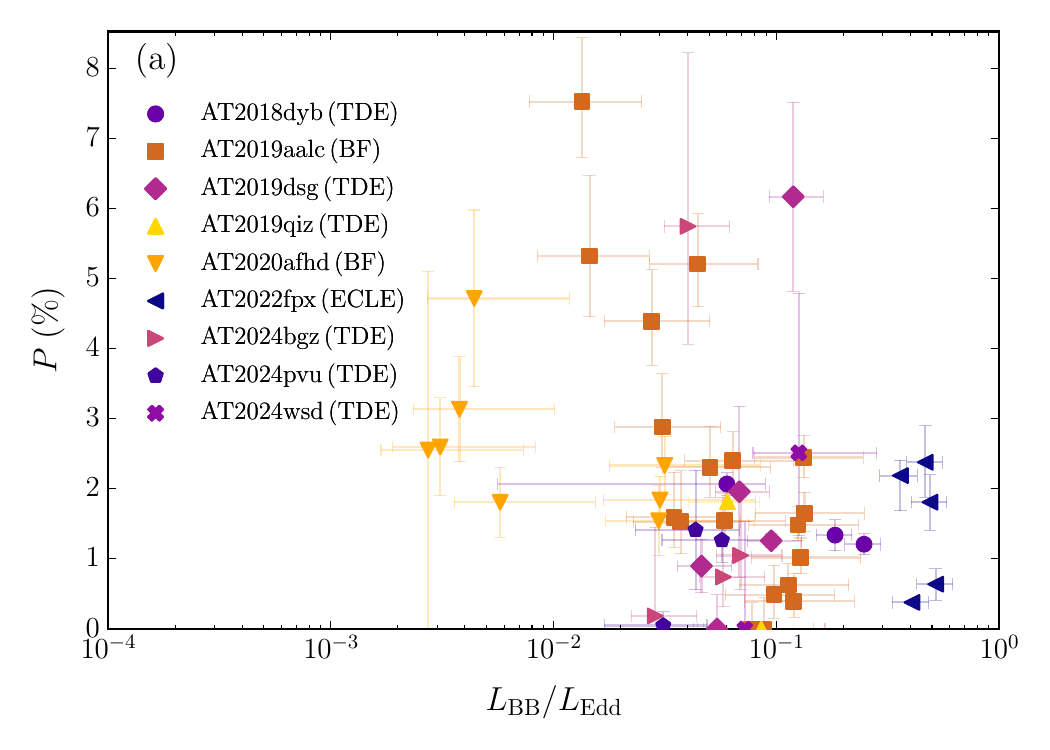}
\includegraphics[width=0.45\textwidth]{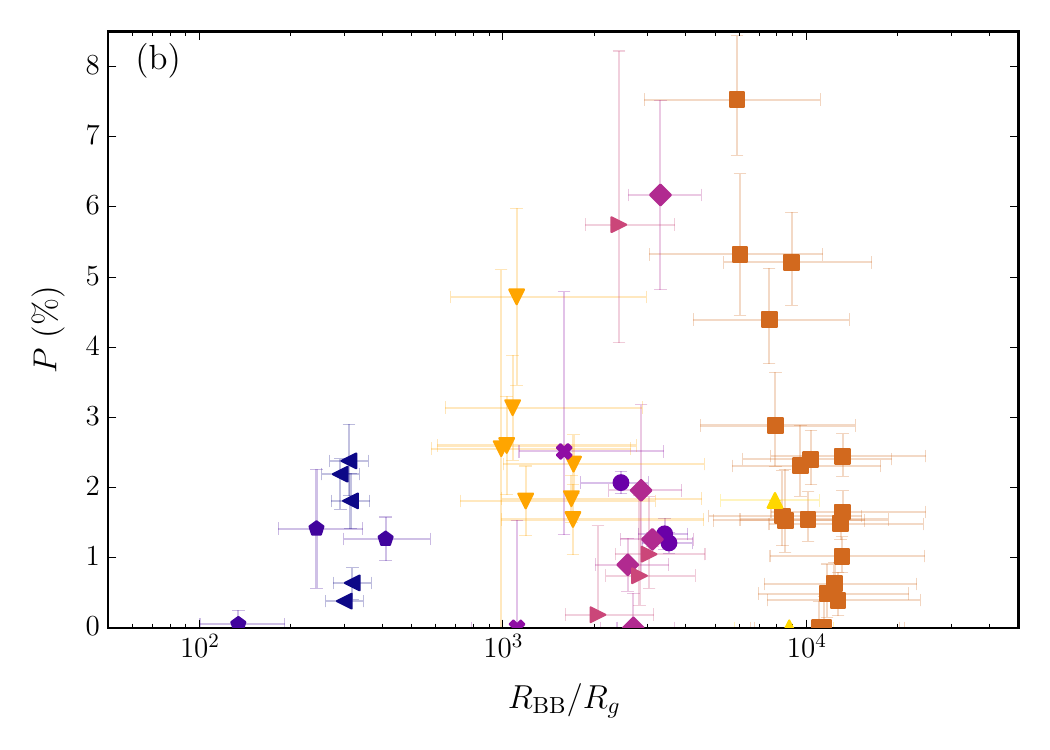}
\end{center}
\caption{Polarization measurements of thermal TDEs, Bowen flares (BF) and AT2022fpx (ECLE) as a function of (a) the Eddington-scaled blackbody luminosity and (b) photospheric radius scaled by gravitational radius, both evaluated at the time of polarimetric observations using MOSFiT blackbody fits to ZTF photometry.}
\label{fig:P_corr_part1}
\end{figure*}

\begin{figure}
\begin{center}
\includegraphics[width=0.5\textwidth]{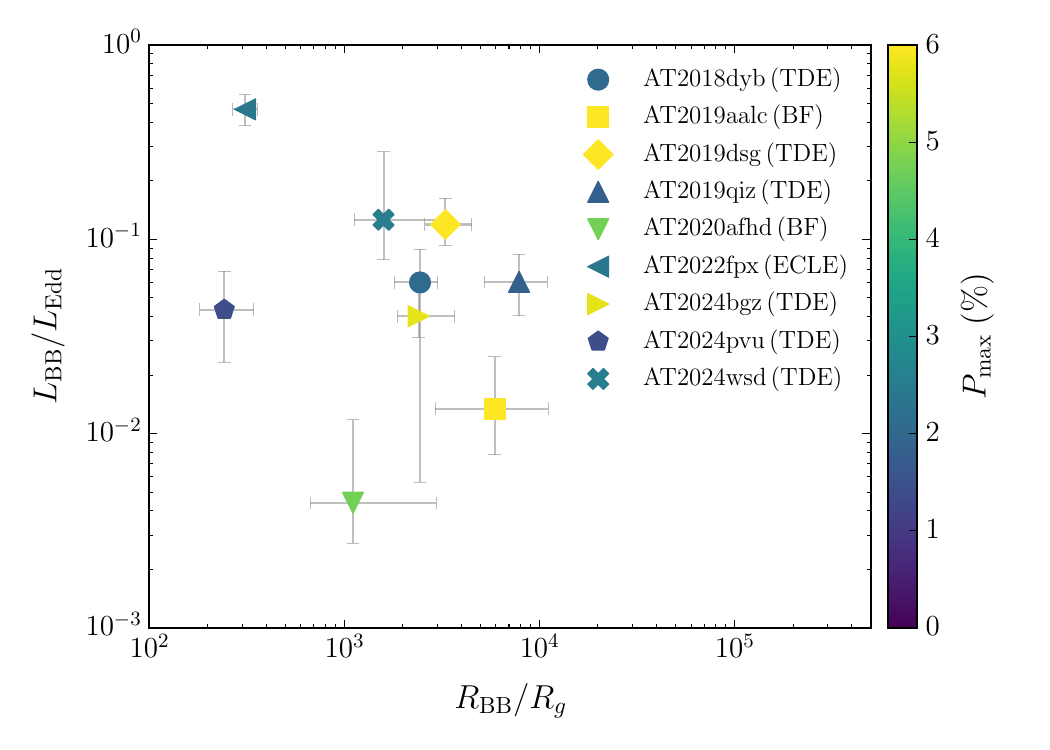}
\end{center}
\caption{Eddington-scaled blackbody luminosity of thermal TDEs, Bowen flares (BF), and AT2022fpx (ECLE) as a function of photospheric radius normalized by gravitational radius, both evaluated at the time of maximum polarization ($P_\mathrm{max}$) of each event. Marker colour indicates the maximum polarization degree measured for each event.}
\label{fig:P_corr_part2}
\end{figure}

Figures \ref{fig:P_corr_part1} and \ref{fig:P_corr_part2} show  that moderate polarization occurs at later times for nuclear transients with low Eddington ratios and highly extended photospheres. This suggests, as we have argued, that as the level of accretion decreases, we can expect to see more asymmetric reprocessing layers from different viewing angles \citep{2022ApJ...937L..28T}. Since the density and velocity of the outflow are highly dependent on the inclination angle \citep{2018ApJ...859L..20D}, TDEs with low Eddington ratios and highly extended photospheres are likely to show varying levels of polarization (Figures \ref{fig:P_corr_part1} and \ref{fig:P_corr_part2})

In this context, the initial low polarization degree of the AT2024pvu ($P\approx 1.3 \%$) and its subsequent decline are consistent with several reprocessing scenarios: outflows from an extended low-mass (high-mass) disk viewed nearly edge-on (face-on; \citealt{2022NatAs...6.1193L}), and a low-mass outflow produced by the self-intersection of stellar debris \citep{2023A&A...670A.150C}. The polarization measurements of the TDE-H+He AT2024wsd were acquired during an optically thin regime (see Figure \ref{fig:Pol_restframe_TDEs}) and are also consistent with reprocessing by outflows originating either from the accretion disk itself or the self-intersecting stellar streams. For the TDE-H+He AT2024bgz, a low polarization level ($P\approx 0.9 \%$) was measured near the peak of the optical light curve, during the receding phase of the photosphere and in an already optically thin regime (see Figure \ref{fig:Pol_restframe_TDEs}). A $2\sigma$ increase to $P \approx 6\%$ about 30 days later hints at an asymmetry in the inner photospheric layers,  supporting the model proposed by \citet{2022ApJ...937L..28T}. Overall, the mild polarization levels and stable polarization angles of the three TDEs support reprocessing frameworks involving either disk-driven outflows \citep{2022NatAs...6.1193L} or collision-induced outflows \citep{2023A&A...670A.150C}. Among the seven thermal-dominated TDEs with time-series polarization data ---AT2018dyb, AT2019qiz, AT2019dsg, AT2020mot, and the three TDEs presented here---, we note that only AT2020mot exhibits polarization signatures attributed to a shock-powered, non-thermal component \citep{2023Sci...380..656L}.

\section{Bowen flares: AT2019aalc and AT2020afhd}
\label{sec:BowenFlares}

Here, we discuss the origin of the two Bowen flares, AT2019aalc and AT2020afhd (see Section \ref{sec:Bowenflares_asTDEs}), and their polarization time series (see Section \ref{sec:Bowenflares_pol}).

\subsection{Bowen flares as stellar-fed SMBHs}  \label{sec:Bowenflares_asTDEs}

Bowen flares, such as AT2019aalc \citep{2024arXiv240817419V,2025arXiv250500083S} and AT2020afhd \citep{2024TNSAN..53....1A}, are TDE-like outbursts of intense UV/optical emission that have been detected in AGN. Bowen flares share spectral properties with AGN and TDEs \citep{2019NatAs...3..242T}; they display emission lines that are typical of unobscured AGN, as well as distinct Bowen emission lines that are persistent across flaring time (i.e. He II $\lambda4686$, N III $\lambda4640$, and O III $\lambda3133$). Bowen fluorescence emission lines are very rare in AGN \citep{1990ApJ...362...74S} and, although generally broader, they have been detected in some TDEs of the TDE-H-He spectral class (e.g. \citealt{2019ApJ...873...92B,2023ApJ...942....9H}). Bowen fluorescence requires strong EUV emission to initiate a resonance cascade with He II Ly‐$\alpha$ that ionizes the high-density metal-rich material (oxygen and nitrogen; \citealt{1934PASP...46..146B,1935ApJ....81....1B}) in the vicinity of the SMBH. This suggests a similar mechanism driving TDEs and Bowen flares, which does not get triggered during regular AGN activity.

\subsubsection{Recurring outbursts from partial TDEs}
\label{sec:BF_recurrent}

\begin{figure}
\begin{center}
\includegraphics[width=0.5\textwidth]{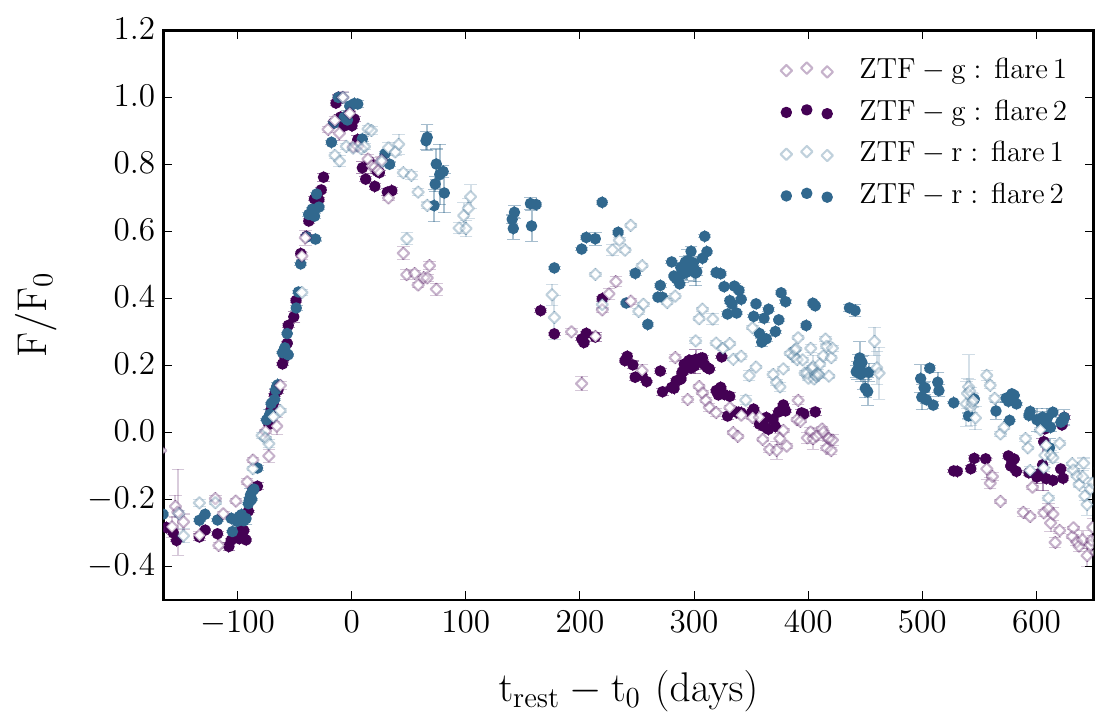}
\end{center}
\caption{Rest-frame ZTF $g$ and $r$ light curves of the two flares of AT2019aalc, binned to 1 day, time-shifted to the optical peak ($t_0$), and normalized to their respective peak fluxes ($F_0$).}
\label{fig:LC_AT2019aalc_profile}
\end{figure}

\begin{figure*}
\begin{center}
\includegraphics[width=0.435\textwidth]{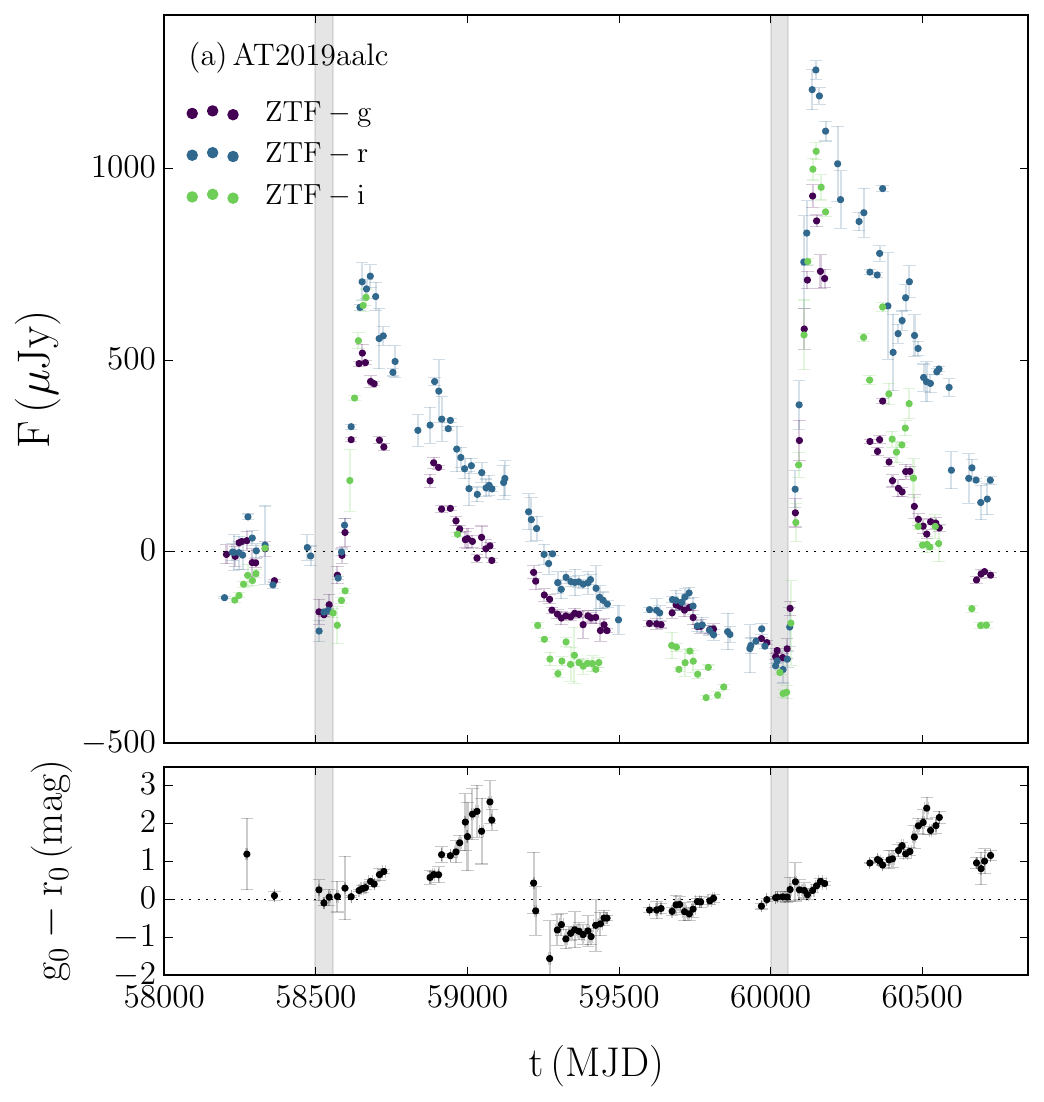}
\includegraphics[width=0.45\textwidth]{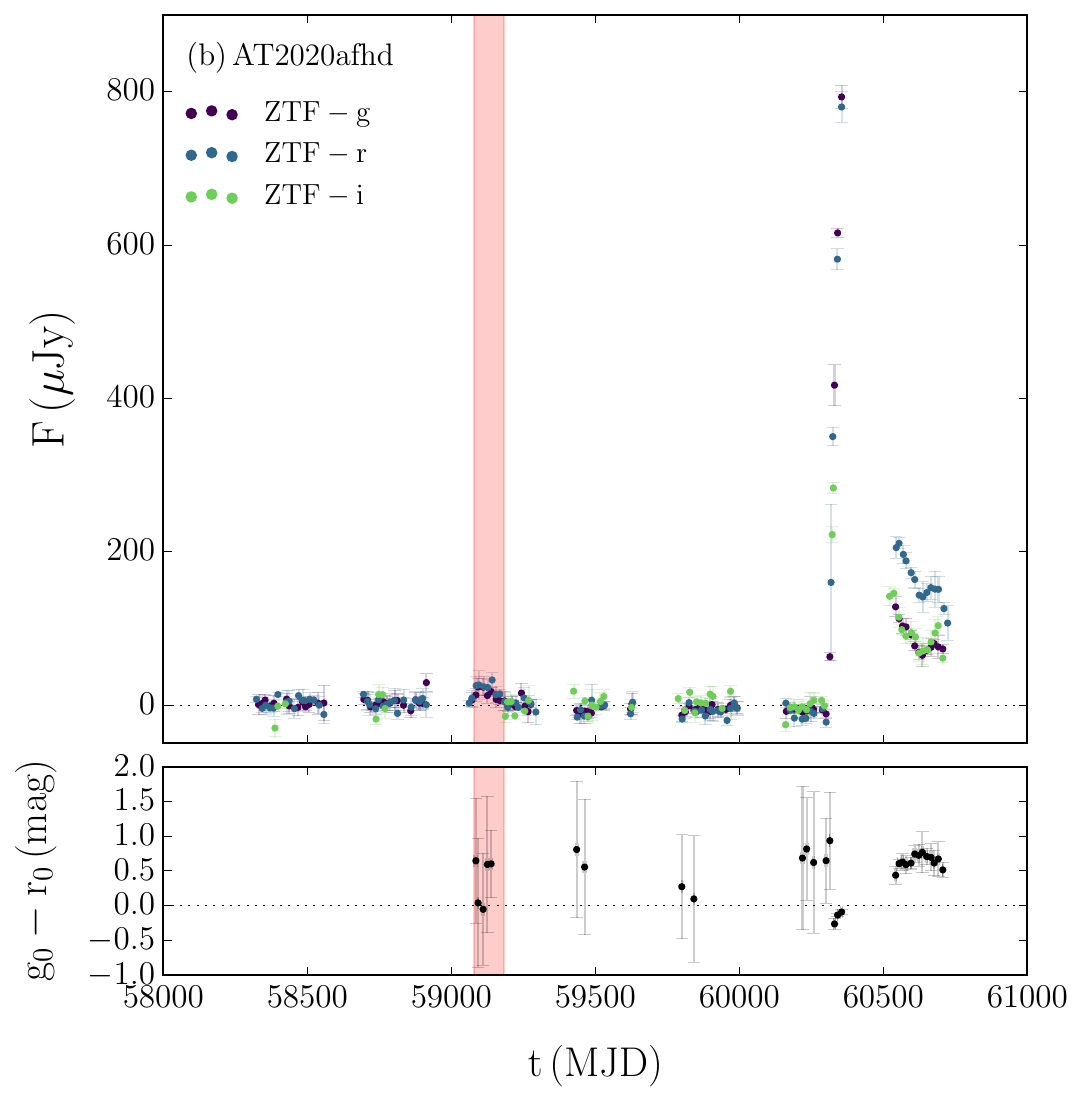}
\end{center}
\caption{ZTF light curves of the repeating (a) AT2019aalc and (b) AT2020afhd, binned to $14 \,$days. The bottom panel displays the colour evolution corrected for Galactic extinction and host galaxy contributions. For (a) AT2019aalc, the shaded grey regions indicate light curve segments corresponding to flux dips before the flares identified by Bayesian blocks at significance levels of $34\sigma$ (first) and $8\sigma$ (second). For (b) AT2020afhd, the $30\sigma$ flux excess is shown in red.}
\label{fig:LC_repeating}
\end{figure*}

Bowen flares have been proposed as a new form of AGN activity \citep{2019NatAs...3..242T}. However, we argue that the recurrence of these flares strongly favours a repeating partial TDE scenario \citep{2013ApJ...777..133M,2023ApJ...944..184L,2025ApJ...979...40L,2024ApJ...974...80B} rather than stochastic (e.g. \citealt{2009ApJ...698..895K,2010ApJ...721.1014M}) or extreme AGN variability \citep{2019ApJ...883...94T}. That is, AT2019aalc exhibited two flares, separated by 4.1 years, with a consistent spectral and temporal profile across cycles (see Figures \ref{fig:LC_AT2019aalc_profile}-\ref{fig:LC_repeating}, a). Using Bayesian blocks on the ZTF $r$-band light curve \citep{2013ApJ...764..167S,2025ApJ...979...40L}, we also identified a $30\sigma$ emission excess about 3.4 years before the main flare of AT2020afhd (see Figure \ref{fig:LC_repeating}). The second flare of AT2019aalc and AT2020afhd was approximately 2 and 36 times brighter than the first, respectively. This behaviour is consistent with a low-mass star being tidally heated and gradually stripped of mass over multiple passages around the SMBH ---initially producing weak flares that exponentially intensify in the final encounters before complete disruption \citep{2025ApJ...979...40L}.

MOSFiT modelling of AT2019aalc and AT2020afhd suggests that they are TDEs of low-mass stars, with mass $\approx 0.2 M_{\odot}$ during the last passages (see Table \ref{tab:mosfit}). The compact photospheres of the two Bowen flares and $\approx 7 \times 10^6 M_{\odot}$ SMBHs are consistent with the average properties of the TDE-H-He spectral class \citep{2023ApJ...942....9H}. For AT2019aalc, we find that the viscous timescales of both flares are surprisingly long ($\approx 70\,$days), which bottlenecks fallback accretion and results in an optical light curve with a shallower decay and flatter peak (see Figure \ref{fig:LC_abs}; \citealt{2015ApJ...809..166G,2019ApJ...872..151M}). The viscous delay of AT2019aalc is consistent with the timescales of a flux reduction in the optical bands that occur before each flare. That is, the Bayesian blocks analysis reveals that there was a drop in brightness of 61 days before the first flare at a significance of $34\sigma$ and a $8\sigma$-emission dip that lasted 58 days before the second flare (see Figure \ref{fig:LC_repeating}-a). Their lack of colour change is consistent with grey obscuration by the temporary suspension of dust-free, optically-thick stellar debris.

\subsubsection{AT2019aalc: post-flare reconfiguration of the inner disk}

The Seyfert 1 polar view of AT2019aalc provides an unobscured view of its nucleus and accretion disk. Following its first flare, the optical continuum of the host dimmed by $ \approx 0.2 \,$mag relative to its pre-burst level ---a behaviour reminiscent of CSS100217 \citep{2017ApJ...843L..19M}. This decline likely reflects a significant reconfiguration of the AGN accretion disk, driven by the interaction between the disrupted stellar debris and the pre-existing disk. Such interaction can generate shocks that efficiently transport angular momentum \citep{2025arXiv250619900D}, triggering a rapid inward inflow and leading to a partial depletion of the inner disk \citep{2019ApJ...881..113C}. Although the incoming stream might partially replenish the inner disk, the timescales for full recovery are uncertain. Since part of the optical emission in AGN is enhanced through the reprocessing of the inner disk \citep{1991ApJ...371..541K}, any reduction of the inner disk luminosity would naturally lead to a decline in the observed optical brightness. Moreover, the blue colours observed immediately after the first flare ($\approx 59250 \,$MJD) are consistent with a TDE that shock-heated the inner disk; as the disk cools, its continuum reverts to the pre-flare redder state ($\approx 59250-60000 \,$MJD; see Figure \ref{fig:LC_repeating}-a). Alternatively, if the TDE induced a geometrical reconfiguration of the inner disk (e.g., a tilted disk, warping), the reprocessing efficiency of the X-ray emission by the outer disk can also be altered \citep{2009MNRAS.394..427B}, resulting in the observed optical dimming.

\begin{figure*}
\begin{center}
\includegraphics[width=\textwidth]{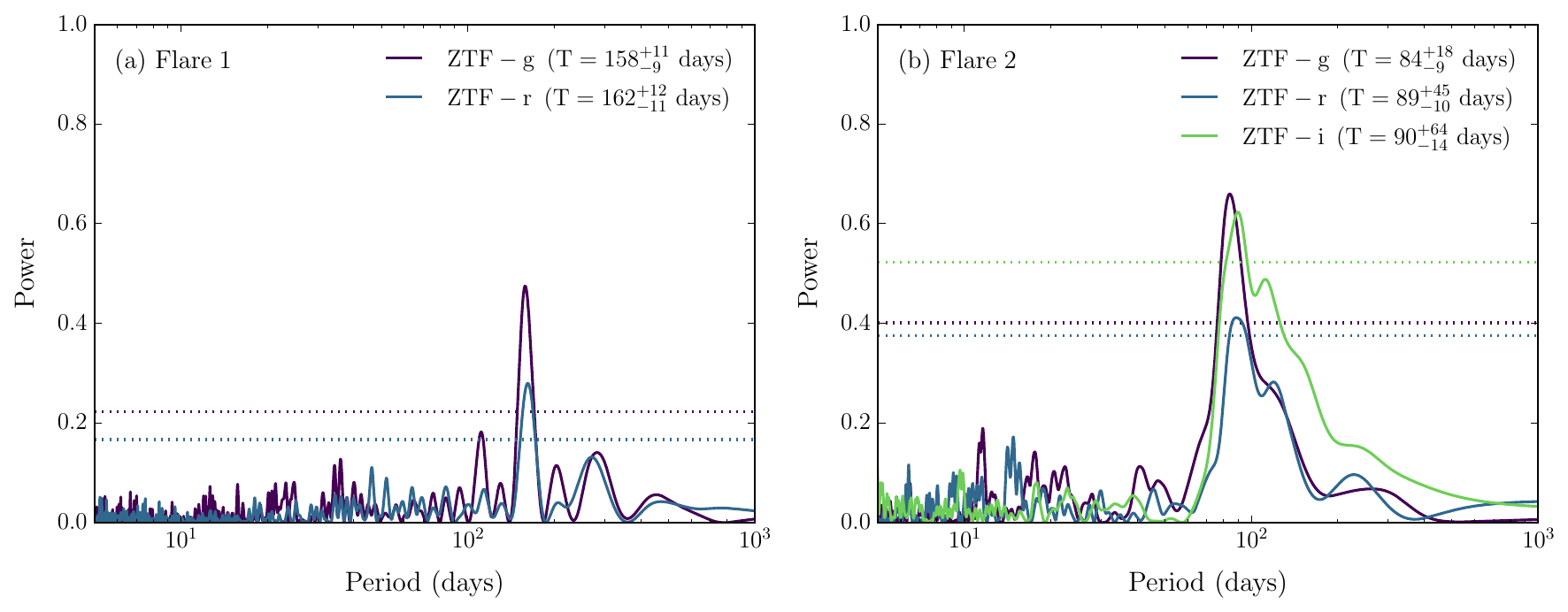}
\end{center}
\caption{Lomb-Scargle periodograms of AT2019aalc from ZTF forced photometry, binned to 1-day intervals. Panel (a) shows the periodograms for the first flare of AT2019aalc, using data from the time of its peak up to the onset of the next flare ($58700-59900 \,$MJD). To account for the underlying fading of the continuum beyond the first flare, we subtracted a 4th-degree polynomial fit. Panel (b) displays the periodograms of the second flare of AT2019aalc ($60200-60730 \,$MJD). In this case, a 2nd-degree polynomial was subtracted to remove the continuum. Uncertainties of the period (T) correspond to the FWHM of the detected peaks. The dotted horizontal lines represent the $3\sigma$ false-alarm probability thresholds, estimated via bootstrapping. Note that none of the peaks in the $i$-band light curve of the first flare exceeded $3\sigma$ and were thus excluded.}
\label{fig:LC_AT2019aalc_peridogram}
\end{figure*}

\subsubsection{AT2019aalc: quasi-periodic oscillations from a precessing inner disk} \label{sec:QPOs}

AT2019aalc displayed repeated flux excesses superimposed on the overall emission decay. We subtracted the smooth, decreasing baseline with a polynomial fit and performed a Lomb-Scargle periodogram analysis \citep{1982ApJ...263..835S}. With false-alarm probabilities estimated via bootstrap simulations, we identified two sets of quasi-periodic oscillations (QPOs) across ZTF bands at more than $3\sigma$ significance (see Figure \ref{fig:LC_AT2019aalc_peridogram}). The first QPO has a period of $T \approx 162 \,$days and is observed from the onset of the first flare until the start of the second flare, while the second set of QPO with a period of $T \approx 88 \,$days is detected during the second flare. In a TDE scenario, QPOs could arise from modulations in the reprocessing layer or intrinsic variability of the inner accretion disk.

If these QPOs reflect the dynamical timescales of the orbiting material, then assuming circular Keplerian orbits with timescales $t_{\rm kep} = 2\pi \sqrt{r^3/ G M_{\rm BH}}$, the material is orbiting at $r \approx 10 ^{15}$ cm. This is consistent with the size of TDE reprocessing photospheres \citep{2022MNRAS.515.5604N,2023ApJ...942....9H} as well as the optical-emitting region of an AGN disk \citep{1973A&A....24..337S}. In this context, QPOs could be due to inhomogeneities in the reprocessing layer from overdensities in the circularization of the TDE debris or material ejected in debris-disk collisions \citep{2021ApJ...914..107C}, which would periodically change the covering factor. The higher-frequency QPO of the second flare would imply that the material responsible for the QPOs could be orbiting closer to the SMBH.

However, the QPO signal remained once the first flare of AT2019aalc faded beyond pre-flare continuum levels (see Figure \ref{fig:LC_repeating}-a), which argues against overdensities from the TDE optically-thick reprocessing layer. Instead, we suggest that the TDE caused a long-lived change in the AGN inner disk \citep{2019ApJ...881..113C}. Given the short lag timescales ($\approx 0.7 \,$days) between the inner disk and the TDE reprocessing layer (or optical-emitting AGN disk region), the observed QPOs could arise from geometric changes in the inner disk, which would modulate what is ultimately reprocessed at larger radii. This interpretation is further supported by contemporaneous X-ray observations during the optical flux excesses, which revealed a tenfold increase in X-ray flux \citep{2024arXiv240817419V,2025arXiv250500083S}. Such behaviour is consistent with a scenario where the inner accretion disk is precessing over time. In particular, frame-dragging precession of the accretion disk can be induced if the infalling stellar debris is injected misaligned with the SMBH spin axis \citep{2012PhRvL.108f1302S,2024Natur.630..325P}. Assuming a circularization radius of $R_{\rm c}=2R_{\rm t}$, where $R_{\rm t}= R_{\star} (M_{\rm BH}/M_{\star})^{1/3}$ is the tidal radius, the Lense-Thirring torque timescale for the solid-body precession of a narrow ring is $t_{\rm LT} =(\pi c^3 R_{\rm c}^3)/(a G^2 M_{\rm BH}^2)$ \citep{1918PhyZ...19..156L,1998ApJ...492L..59S}. For the range of allowed masses of the disrupted star (see Table \ref{tab:mosfit}) and the period of the QPOs (see Figure \ref{fig:LC_AT2019aalc_peridogram}-b), this implies a slowly-rotating SMBH with spin ${\mid a \mid} \lesssim 0.07$. In this context, the higher-frequency QPO of the second flare could be caused by a smaller circularization radius from a deeper star-SMBH encounter. Given the evidence of persistent TDE disks years after the optical flares \citep{2019ApJ...878...82V}, the intermittent view of the inner disk and low SMBH spin would modulate the reprocessed optical emission until the alignment of the disk with the SMBH spin, years later \citep{2016MNRAS.455.1946F}.


\subsection{Polarization of Bowen flares} \label{sec:Bowenflares_pol}

Here, we discuss the interpretation of the polarization time series of AT2019aalc (Sections \ref{sec:AT2019aalc_early}-\ref{sec:AT2019aalc_late}) and AT2020afhd (Section \ref{sec:AT2020afhd_pol}).

\subsubsection{AT2019aalc: early-time asymmetries in the accretion outflows} \label{sec:AT2019aalc_early}

Between 30 days pre-optical peak and 25 days post-peak, the polarization degree of AT2019aalc rose from $P \approx 0\%$ to $P \approx 2\%$. This trend closely resembles that of AT2019qiz, a TDE-H+He. AT2019qiz was initially unpolarized near optical-peak time but exhibited a polarization increase to $P \approx 1\%$ around 29 days post-peak \citep{2022MNRAS.515..138P} ---a value likely underestimated due to host galaxy dilution, and potentially as high as $P \approx 2\%$ when accounting for it (see Figure \ref{fig:Pol_restframe_TDEs}). To explain the increase of the polarization degree with time and the coincident rise of X-ray emission, two mechanisms were proposed: a receding quasi-spherical photosphere that exposed a growing asymmetry deep within the outflow \citep{2022MNRAS.515..138P} or a strong decrease of an initially-high optical depth coupled with an intrinsically asymmetric outflow \citep{2023A&A...670A.150C}. For AT2019aalc, the increasing polarization degree and stable polarization angle were observed during the optically-thick regime (see Figure \ref{fig:Pol_restframe_TDEs}), which is consistent with disk-driven emission reprocessed by large-scale, asymmetric outflows \citep{2022NatAs...6.1193L,2023A&A...670A.150C}. In this context, the polarization rise of AT2019aalc indicates that the outer layers of the photosphere began to thin around $\approx 60115$ MJD (i.e., at $\approx 20$ days before the optical peak), allowing single‐scattered photons to start escaping. This is consistent with the contemporaneous X-ray upturn, with a peak at $\approx 60131$ MJD \citep{2024arXiv240817419V,2025arXiv250500083S}. Although the initial X-ray emission peak of AT2019aalc was at least one magnitude brighter than that of AGN at comparable redshift \citep{2018ApJ...852...37A}, the X-ray-to-optical luminosity ratio was low ($L_{\rm X} / L_{\rm opt}  \approx 10^{-2}$). This low ratio, combined with the high-density environment necessary for Bowen fluorescence, further supports a reprocessing-dominated TDE scenario \citep{2016ApJ...827....3R}.

The absence of a clear trend in the polarization angle of AT2019aalc before 25 days post-optical peak is indicative of an axisymmetric photosphere during this early optically thick phase. However, mild stochastic variability was detected in both polarization degree and angle ($\Delta P \approx  0.4 \%, \Delta \theta \approx 15 \degr $), likely due to a clumpy, inhomogeneous medium ---similar to what has been proposed for AT2019qiz \citep{2022MNRAS.515..138P} and AT2022fpx \citep{2024arXiv240304877K}. AT2019aalc, AT2019qiz, and AT2022fpx showed coronal lines, in particular, high-ionization iron emission lines \citep{2023MNRAS.525.1568S,2024arXiv240304877K,2024arXiv240817419V}. The presence of such lines is consistent with the polarization variability, as they are thought to arise from the leakage of a hard continuum ---through, e.g., clumpy medium near the SMBH---, which subsequently ionizes pre-existing material at larger distances \citep{2023MNRAS.525.1568S}.

\subsubsection{AT2019aalc: late-time evidence of a tilted disk geometry} \label{sec:AT2019aalc_late}

As shown in Section \ref{sec:QPOs}, the second flare of AT2019aalc exhibits a QPO with a period of $T \approx 90\,$days, which we have interpreted as disk precession. Phase-folding the polarization measurements taken beyond $ \approx 190 \,$days post-peak reveals this modulation in the polarization signal as well. That is, the polarization angle presents a modulation with amplitude $\Delta \theta \approx 40\,\degr$, peaking in phase with the flux maxima (see Figure \ref{fig:Pol_LC}-a). The polarization degree tends to decrease during the rebrightenings, likely due to the partial cancellation of the polarization vectors. Altogether, these observations further suggest that by $\approx 190\,$days post-optical peak, the TDE material had already settled into an accretion disk, misaligned with the SMBH spin axis.

\subsubsection{AT2020afhd: evidence of direct and scattered emission} \label{sec:AT2020afhd_pol}

To date, polarimetric monitoring of changing‑look AGN at optical wavelengths has shown that their dramatic spectral transitions are likely driven by intrinsic variations in the accretion rate, rather than changes in dust obscuration along the line of sight \citep{2017A&A...607A..40M,2017A&A...604L...3H,2019A&A...625A..54H}. In a canonical AGN type 2, the central engine is obscured behind a dusty torus and a UV/optical flare such as that of AT2020afhd should only be seen via light scattered from the bi-conical polar outflows into our line of sight. In such configuration, purely scattered light should yield a constant polarization angle \citep{2020A&A...636A..23M}. However, AT2020afhd displays a striking $\Delta \theta = 83 \pm 8 \,\degr$ rotation of the polarization angle during its second flare. This behaviour can be naturally explained if both the direct continuum from the flare and the orthogonally polarized light echo ---scattered by the polar outflows--- are observed \citep{2020A&A...636A..23M}. When the flare is bright, the direct component dominates and, as it fades, the scattered echo causes the net angle to rotate by $\Delta \theta = 90 \,\degr$. This two‑component model for polarization suggests an intermediate-viewing angle of the AT2020afhd system \citep{2020A&A...636A..23M}. We note that a $\Delta \theta \approx 90 \,\degr$ was recently reported during the nuclear flare of AT2023clx, in a LINER galaxy, and was interpreted as the interplay between direct outflow emission and an orthogonally-polarized scattered dust echo \citep{2025arXiv250319024U}.

The detection of direct UV/optical continuum emission through what is normally a Compton-thick line of sight has previously pointed towards an exceptionally powerful, transient increase of the AGN accretion rate in the type-2 changing-look 1ES 1927+654 \citep{2019ApJ...883...94T}. For AT2020afhd, the polarization angle shift is consistent with a temporary unobscured view of the flare, in line with the low hydrogen column density inferred by MOSFiT (see Table \ref{tab:mosfit}). Furthermore, the progressively brighter flares of AT2020afhd point towards a partial TDE as the most plausible driver of the enhanced accretion episode (see Section \ref{sec:BF_recurrent} and Figure \ref{fig:LC_repeating}-b; \citealt{2024ApJ...962L...7W,2025ApJ...979...40L}).

\section{Conclusions} \label{sec:conclusions}

The optical polarimetric campaign presented here provides strong evidence that most thermal TDEs are primarily powered by reprocessing of the emission of newly-formed accretion disks, rather than shocks from stellar-debris streams. The three TDEs from quiescent galaxies (AT2024bgz, AT2024pvu, and AT2024wsd) exhibited mild, slowly-declining degrees of optical polarization ($P \lesssim 6 \%$), with fixed electric-vector position angles. This behaviour is in line with what axisymmetric reprocessing geometries predict \citep{2022NatAs...6.1193L,2023A&A...670A.150C}, and it has now been observed in six of the seven thermal TDEs for which time-resolved polarimetry exists.

The Bowen-flare events of this work are best explained as partial TDEs in AGN rather than a new form of AGN variability. Several independent strands of evidence converge on this conclusion. (i) AT2019aalc has produced two optical outbursts with identical rise and decay profiles ---behaviour naturally expected if a star is repeatedly skimmed and stripped on successive pericenter passages. (ii) In both AT2019aalc and AT2020afhd, the second flare is more luminous than the first, matching the prediction that a partially-disrupted star sheds mass exponentially faster in its final encounters with the SMBH \citep{2025ApJ...979...40L}. (iii) QPOs detected since the onset of the first flare of AT2019aalc ---together with a phase-locked, $\Delta \theta \approx 40 \,\degr$ rotation of the linear-polarization angle--- point to a precessing, tilted inner disk formed by the misaligned interaction of the stellar debris with the SMBH. Additionally, the polarization data of AT2019aalc suggests asymmetric, clumpy outflows ---in line with other coronal-line emitters. By contrast, the $\Delta \theta = 83 \pm 8 \,\degr$ swing in the polarization angle observed in the AGN type-2 AT2020afhd signals a combination of direct and polar-scattered light, and hence an episode of extreme accretion seen by a mid-latitude observer. Taken together, these findings suggest that TDEs occurring in AGN may systematically differ from those in quiescent galaxies, and provide the EUV photons and the dense, optically-thick environment necessary for sustained Bowen fluorescence. Additionally, such events can induce long-lasting changes in the AGN inner flow, thereby imprinting distinctive, time-dependent polarization signatures.

The Vera Rubin Observatory’s photometric survey is expected to raise the annual TDE discovery rate from a few tens to several thousand \citep{2020ApJ...890...73B}, enabling population-scale studies across TDE spectral subclasses. Deep, early-time polarimetric observations with sub-month cadence ---paired with soft X-ray monitoring--- of even a small fraction of these TDEs would allow the systematic mapping of the reprocessing-layer geometries and help discriminate between disk-reprocessing and stream-shock emission models.

\begin{acknowledgements}
N.J-M. acknowledges support from the Alexander von Humboldt Foundation. A.F. and and P.M.V. acknowledge the support from the DFG via the Collaborative Research Center SFB1491 Cosmic Interacting Matters - From Source to Signal.  E.R.-R.  thanks the Heising-Simons Foundation and NSF: AST 1852393, AST 2150255 and AST 2206243. The Liverpool Telescope is operated on the island of La Palma by Liverpool John Moores University in the Spanish Observatorio del Roque de los Muchachos of the Instituto de Astrofisica de Canarias with financial support from the UK Science and Technology Facilities Council.  This project has received funding from the European Union’s Horizon 2020 research and innovation programme under grant agreement No 101004719. Based on observations obtained with the Samuel Oschin Telescope 48-inch and the 60-inch Telescope at the Palomar Observatory as part of the Zwicky Transient Facility project. ZTF is supported by the National Science Foundation under Grant No. AST-2034437 and a collaboration including Caltech, IPAC, the Weizmann Institute for Science, the Oskar Klein Center at Stockholm University, the University of Maryland, Deutsches Elektronen-Synchrotron and Humboldt University, the TANGO Consortium of Taiwan, the University of Wisconsin at Milwaukee, Trinity College Dublin, Lawrence Livermore National Laboratories, and IN2P3, France. Operations are conducted by COO, IPAC, and UW. This work has made use of data from the Asteroid Terrestrial-impact Last Alert System (ATLAS) project. The Asteroid Terrestrial-impact Last Alert System (ATLAS) project is primarily funded to search for near-Earth asteroids through NASA grants NN12AR55G, 80NSSC18K0284, and 80NSSC18K1575; byproducts of the NEO search include images and catalogs from the survey area. This work was partially funded by Kepler/K2 grant J1944/80NSSC19K0112 and HST GO-15889, and STFC grants ST/T000198/1 and ST/S006109/1. The ATLAS science products have been made possible through the contributions of the University of Hawaii Institute for Astronomy, the Queen’s University Belfast, the Space Telescope Science Institute, the South African Astronomical Observatory, and The Millennium Institute of Astrophysics (MAS), Chile.
\end{acknowledgements}

\bibliographystyle{aa}
\bibliography{biblio}

\begin{appendix}
\section{Extended Data Reduction} \label{appendix_data_red}

In Appendix \ref{max_SNR}, we explain how we have maximized the signal-to-noise ratio of the transients; in Appendix \ref{pol_inst_all}, we detail the calibration methods that have been employed to correct the target data for the MOPTOP instrumental effects.

\subsection{Maximizing the signal-to-noise ratio} \label{max_SNR}

Each observing sequence with the MOPTOP consists of $N_{\rm rot}$ full rotations of the half-wave plate. Each full rotation yields 16 images per CMOS, which is a total of 32$N_{\rm rot}$ images per observing run. We improved the signal-to-noise ratio of the detections by aligning frames using the astroalign package \citep{2020A&C....3200384B} and combining them. In an initial approach, each 32$N_{\rm rot}$-image dataset was combined into a 32-images dataset (i.e. 16 images per CMOS), from which we extracted 8 pairs of ($q,u$) values following \cite{2020MNRAS.494.4676S}. In a second attempt to maximize the signal-to-noise ratio of the faint targets, we stacked the 32 images until there were only 8 left (i.e. 4 images per CMOS), from which we derived 2 pairs of ($q,u$) values. The average ($q,u$) measurements obtained from the 32- and 8-frame datasets were consistent within $1\sigma$. For the remainder of this work, we adopted the 8-frame approach due to its higher signal-to-noise ratio. For example, during the 14/06/23 observing run of the AT2019aalc (see Table \ref{table:log}), we detected the transient (and host galaxy) at an average signal-to-noise ratio of $\approx 250$ in each of the 32 frames, which was maximized to a signal-to-noise ratio of $\approx 453$ in each of the 8 frames of the final stack (see Table \ref{table:pol_phot_resultats}).

For each observing sequence, we used variable photometric apertures to account for the varying point-spread function of the sources from different atmospheric conditions. We fitted a 2D Gaussian to the sources and estimated the FWHM of the point-spread function. Note that to later be able to estimate the host galaxy depolarization with pre-flare survey data (see Section \ref{sec:depo}), we included most of the host galaxy emission. That is, we used large photometric apertures of 2FWHM radius, corresponding to 99.998\% of the flux of a point-like source.

After correcting the observations for MOPTOP instrumental polarization effects (see Appendix \ref{pol_inst_all}), we checked the stability of the instrument during observations using bright stars in each field of view ($12.4-14.4 \,$mag in the r band). We also tested the accuracy of the polarization measurements using different photometric apertures in the range of $1.5-4 \,$FWHM. For the polarization time series of the transients and bright stars, there were no polarization variations larger than $1\sigma$. The polarization degree of the bright stars presented $\Delta P \approx 0.03-0.15\%$ departure from the mean during MOPTOP observations; therefore, each polarimetric measurement of the transient was individually corrected for these instrumental offsets (note that in this process, we also subtracted the Galactic ISM polarization). The average polarization of these bright stars is reported in Table \ref{table:prop_env}, which gives an estimate of the Galactic ISM polarization in the lines of sight of the transients.

\subsection{Instrumental polarization corrections} \label{pol_inst_all}

Here, we describe the steps that we have followed to correct the target data by MOPTOP instrumental effects. We first subtracted the instrumental offset in Appendix \ref{qu_inst}; we then corrected the data by the instrumental depolarization in Appendix \ref{ellipt_pol} (alternatively, see Appendix \ref{constant_pol}, where we describe a correction assuming a constant depolarization factor); finally, we corrected the polarization angle for instrumental rotation effects in Appendix \ref{angle_offset}.

\subsubsection{Stokes parameters offset} \label{qu_inst}

\begin{figure*}
\begin{center}
\includegraphics[width=\columnwidth]{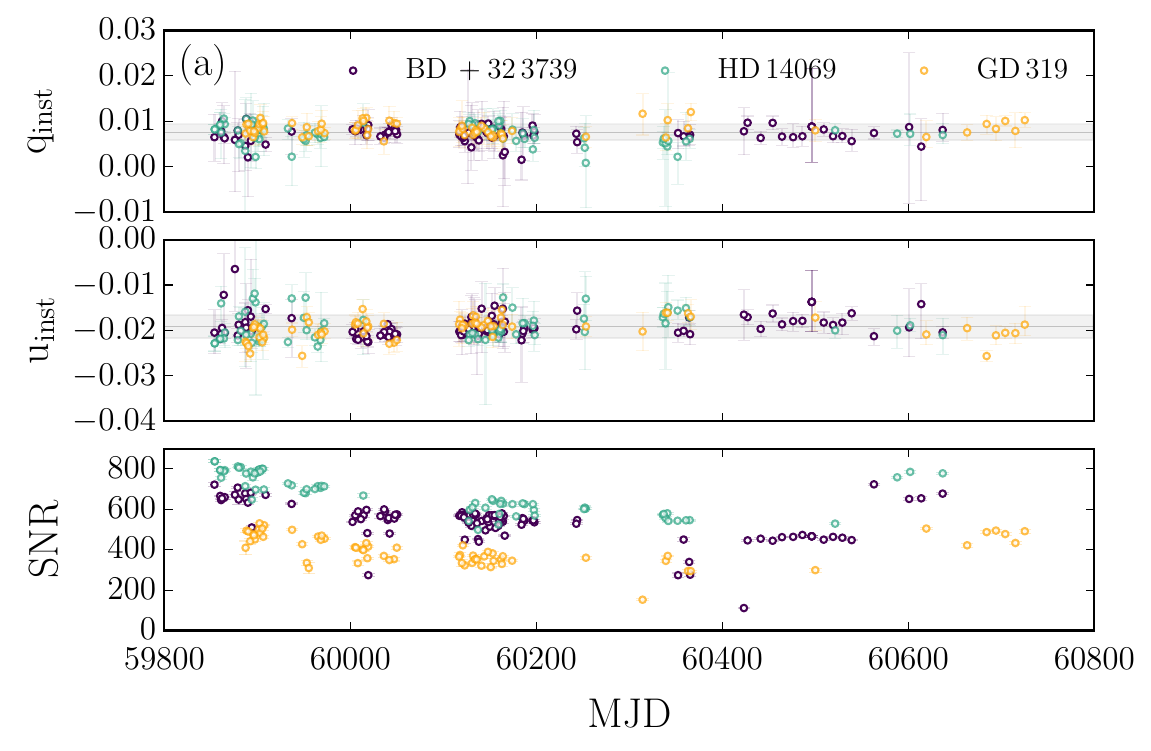}
\includegraphics[width=\columnwidth]{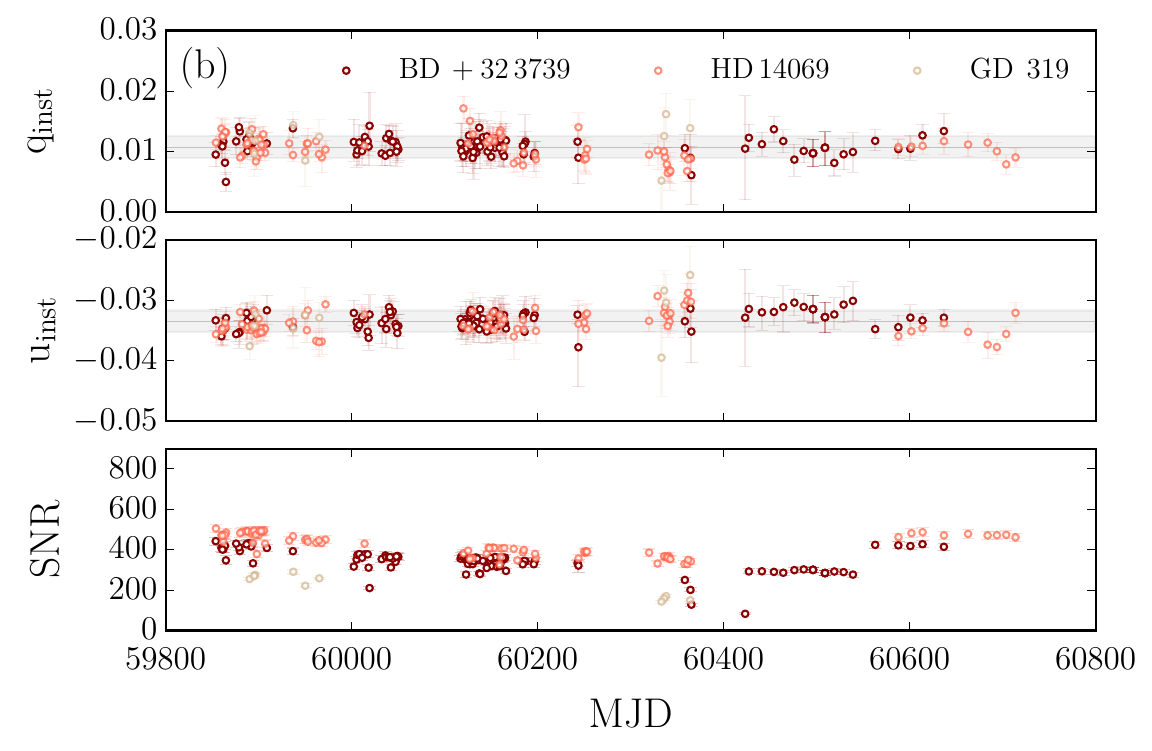}
\end{center}
\caption{Time series of the MOPTOP instrumental zero points ($q_{\rm inst}, u_{\rm inst}$) for the (a) MOP-L and (b) MOP-I filters. The unpolarized stars that have been used for the determination of these zero points are GD 319, BD +32 3739, and HD 14069. The grey lines correspond to the mean instrumental offset and the shaded-grey regions to its standard deviation. The SNR panel corresponds to the target average signal-to-noise ratio.}
\label{fig:Pinstr_timeseries}
\end{figure*}

To correct for the instrumental polarization, we analysed 244 routine MOP-L and 199 MOP-I observations of the unpolarized standards GD 319, BD +32 3739, and HD 14069 \citep{1990AJ.....99.1243T} between dates 01/10/22 and 18/02/25. We calculated the Stokes parameters of the unpolarized standards to determine the instrumental polarization offset ($q_{\rm inst}$, $u_{\rm inst}$; see Figures \ref{fig:Pinstr_timeseries}). The average Stokes parameters\footnote{Note that the average values of instrumental polarization derived here are in agreement with those reported in the Liverpool Telescope website: \url{https://telescope.livjm.ac.uk/TelInst/Inst/MOPTOP/}} were $\bar{q}_{\rm inst, \, L}= 0.008 \pm 0.002 $, $\bar{u}_{\rm inst, \, L}= -0.019 \pm 0.003 $ for the MOP-L filter, and $\bar{q}_{\rm inst, \, I}= 0.011 \pm 0.002 $, $\bar{u}_{\rm inst, \, I }= -0.033 \pm 0.002$ for the MOP-I filter.

Using a Pearson’s correlation test on the time series, we detected the following significant trends with p-values$\, < 0.05$; the signal-to-noise ratio of the standard stars presented a monotonic decrease across years caused by the normal degradation of the Liverpool Telescope primary mirror (with correlation coefficients $|r| > 0.7$). This trend continued until 02/08/24, when the mirror was re-aluminised\footnote{\url{https://telescope.ljmu.ac.uk/News/\#realuminising2024}}, resulting in an increase in signal-to-noise ratio by a factor of $\approx 1.6$. Additionally, the $q_{\rm inst}$ offset was decreasing with time ($|r|_{\lbrace \rm L, I \rbrace}=0.1,0.2$ for the MOP-L and MOP-I filters, respectively), whilst the $u_{\rm inst}$ increased ($|r|_{\lbrace \rm L, I \rbrace } =0.2,0.4$). To account for the observed trends, we corrected each raw science measurement ($q_{\rm raw}$, $u_{\rm raw}$) for an average instrumental offset determined around the same observing time as the science target ($q_{\rm inst}$, $u_{\rm inst}$). The instrumental offsets were then subtracted from the science measurement as $(q^{\prime}, u^{\prime}) = (q_{\rm raw}, u_{\rm raw})-(q_{\rm inst}, u_{\rm inst})$.

\subsubsection{Constant depolarization}
\label{constant_pol}

To calculate the degree of depolarization of the MOPTOP polarimeter with the MOP-L and MOP-I filter setups, we used routine observations of the following polarized standards: BD +64 106, VI Cyg 12, HD 251204, HD 155197, and HILT 960 \citep{1990AJ.....99.1243T,1992AJ....104.1563S,1999AcA....49...59W}. The standards were observed between the dates 01/10/22 and 19/02/25, and we analysed 562 MOP-L and 556 MOP-I observations.

Assuming uniform depolarization across the Stokes plane, the ($q^{\prime}, u^{\prime}$) measurements are expected to lie along characteristic polarization circles, resulting from variations in the telescope’s rotation angle at each observation epoch (see Appendix \ref{angle_offset}). Under this assumption, the catalogue polarization value ($P_{\rm ref}$) and the measured polarization ($P^{\prime} = 100 \sqrt{q^{\prime \, 2} + u^{\prime\, 2}}$) are directly related as $P^{\prime}=P_{\rm ref} \, D_{\rm c}$. We measured depolarization factors $ D_{\rm c, \, L} = 0.87 \pm 0.01 $ and $ D_{\rm c, \, I} = 0.81 \pm 0.02$ for the MOP-L and MOP-I filters, respectively; see Figures \ref{fig:Pdepo}.

\subsubsection{Elliptical depolarization}
\label{ellipt_pol}

Former polarimeters at the Liverpool Telescope presented elliptical depolarization \citep{arnold2017,jordana2021}. MOPTOP also suffers from this instrumental effect and the ($q^{\prime},u^{\prime}$) data are distributed in ellipsoids in the Stokes plane instead of circles (see Figures \ref{fig:Pellipt_MOPL}-\ref{fig:Pellipt_MOPI}, a), which causes an angle dependency on the depolarization. Therefore, next, we characterized the degree of ellipticity of the MOPTOP and corrected the data for it before calculating a new depolarization factor ($D_{\rm \epsilon}$). 

For each MOPTOP filter, we simultaneously fitted an ellipse to the normalized ($q^{\prime}$, $u^{\prime}$) data of the five polarized standards, with parameters the major axis scaling ($a$), the minor axis scaling ($b$), and the ellipse rotation angle ($\psi$). We find the following parameters; for the MOP-L filter, $a_{\rm L} = 1.080 \pm 0.005$, $b_{\rm L}  = 0.919 \pm 0.004$, and $\psi_{\rm L}  = 106.4 \pm  1.3 \, \degr$ (with eccentricity $\epsilon_{\rm L} = 0.53 \pm 0.01$); for the MOP-I filter, 
$a_{\rm I} = 1.080 \pm 0.006$, $b_{\rm I} = 0.941 \pm 0.004$, $\psi_{\rm I} = 108.8 \pm  1.4 \, \degr$ ($\epsilon_{\rm I} = 0.49 \pm 0.01$). Note that the eccentricity levels are similar to those found for the MOPTOP predecessor RINGO3 \citep{jordana2021}. 

\begin{figure*}
\begin{center}
\includegraphics[width=7.5cm]{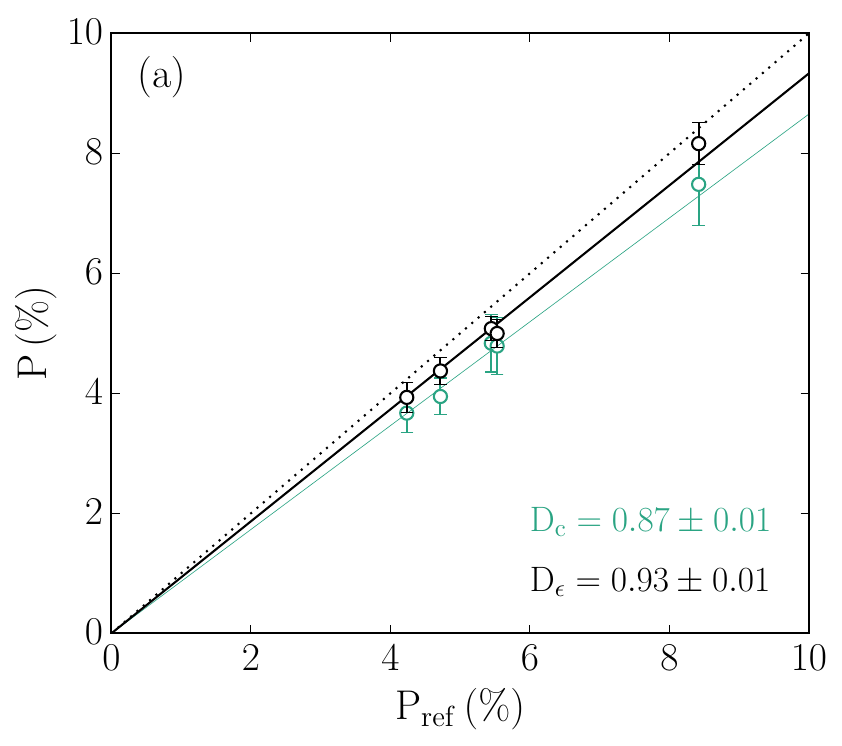}
\includegraphics[width=7.5cm]{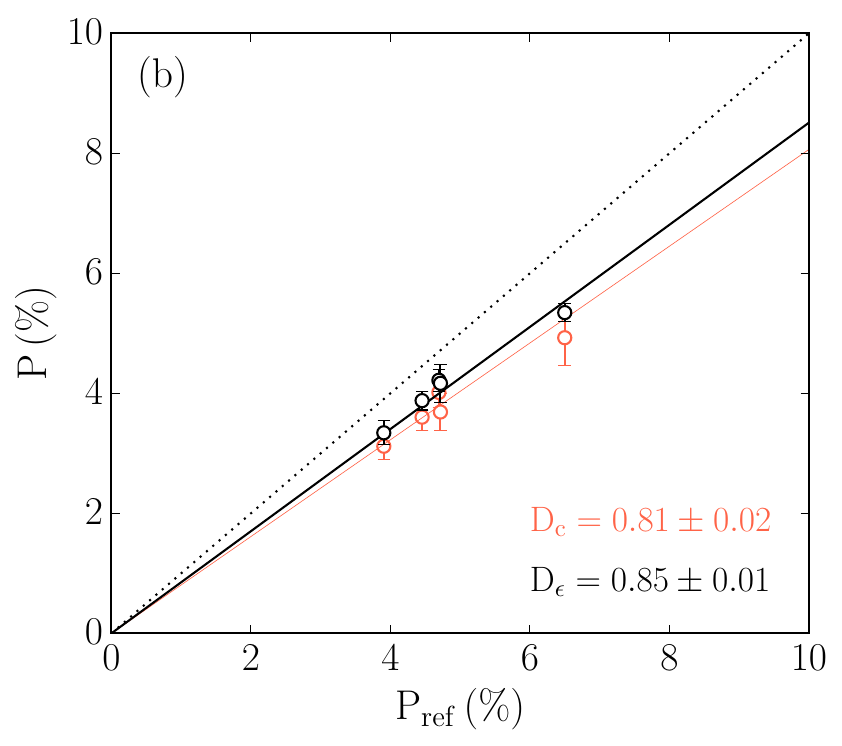}
\end{center}
\caption{MOPTOP instrumental depolarizarion for the (a) MOP-L and (b) MOP-I filters. The $P_{\rm ref}$ is the polarization degree of BD +64 106, VI Cyg 12, HD 251204, HD 155197, and HILT 960 reported in \cite{1990AJ.....99.1243T,1992AJ....104.1563S,1999AcA....49...59W}, and $P$ is that measured with the MOPTOP; the dotted line represents the 1:1 relation. The solid lines are linear fits to the data; i.e. $P = P_{\rm ref} \, D$, where $D$ is the depolarization factor. Coloured-marker data ($D_{\rm c}$) do not include ellipticity corrections and black-marker data ($D_{\epsilon}$) have been corrected for ellipticity.}
\label{fig:Pdepo}
\end{figure*}

Next, we corrected the ellipticity and derived the transformed Stokes parameters  ($q_{\epsilon}^{\prime}, u_{\epsilon}^{\prime}$). Starting from the general equation of an ellipse, we have that each pair of ($q_i^{\prime}$, $u_i^{\prime}$) satisfies
\begin{align}
q_i^{\prime} &= a_i \, \cos(\psi) \, \cos(\phi_i)-b_i\, \sin(\psi) \, \sin(\phi_i) = \nonumber \\
&= a_i \,\cos(\psi) \,\cos(\phi_i)-a_i(1-\delta)\, \sin(\psi) \,\sin(\phi_i) \,, \\
u_i^{\prime} &=a_i \,\sin(\psi) \, \cos(\phi_i)+b_i \,\cos(\psi)\,\sin(\phi_i) = \nonumber \\
&= a_i\,\sin(\psi)\,\cos(\phi_i)+a_i(1-\delta)\,\cos(\psi)\,\sin(\phi_i) \,,
\end{align}where the angle is
\begin{align}
\phi_i &= \tan^{-1} \bigg(\frac{u_i^{\prime}}{q_i^{\prime}} \bigg)- \psi
\end{align} and we have defined $\delta = (a-b)/a$, which takes values $\delta_{\rm L} = 0.149 \pm 0.006$ and $\delta_{\rm I} = 0.129 \pm 0.006$, depending on the MOPTOP filter. We corrected each pair of ($q_i^{\prime}$, $u_i^{\prime}$) values to obtain the ellipticity-corrected Stokes parameters ($q^{\prime} _{\epsilon}$, $u_{\epsilon}^{\prime}$) as follows
\begin{align}
q_{\epsilon}^{\prime} &= q_i^{\prime} - a_i \, \delta \, \sin(\phi_i) \, \sin(\psi) \nonumber \\
u_{\epsilon}^{\prime}  &= u_i^{\prime} + a_i \,  \delta \, \sin(\phi_i) \, \cos(\psi) \, ,
\end{align} which corresponds to a polarization circle with a radius the major axis $a_i$. For each pair of ($q_i^{\prime}$, $u_i^{\prime}$), the parameter $a_i$ was calculated as

\begin{align}
p_i^{\prime} &= \sqrt{a_i^2 \cos^2 (\phi_i) + b_i^2 \sin^2 (\phi_i)} \, , \\
p_i^{\prime \, 2} &= a_i^2 \cos^2(\phi_i)+a_i^2(1-\delta)^2 \sin^2(\phi_i) = \nonumber\\
&= a_i^2 \, \big(\cos^2(\phi_i) + (1 - 2\delta + \delta^2) \sin^2(\phi_i) \big) \, ,\\
a_i &= \sqrt{\frac{q_i ^{\prime 2} + u_i ^{\prime 2}}{\cos^2(\phi_i) + (1-2\delta+\delta^2) \sin^2(\phi_i)}}.
\end{align} After correcting for ellipticity in Figures \ref{fig:Pellipt_MOPL}-\ref{fig:Pellipt_MOPI} (b), we find depolarization factors $ D_{\epsilon, \, \rm L } = 0.93 \pm 0.01$ and $ D_{\epsilon, \, \rm I } = 0.85 \pm 0.01$ for the MOP-L and MOP-I filters, respectively (see Figures \ref{fig:Pdepo}). The Stokes parameters were corrected as $(q^{\prime \prime}, u^{\prime \prime}) = (q^{\prime}_{\rm \epsilon}, u^{\prime}_{\rm \epsilon}) / D_{\epsilon}$.

To evaluate the long-term stability of MOPTOP, we analysed the polarization degree of polarized standard stars over a two-year period. We found that the Pearson correlation coefficients of the polarization time series of the standard stars shown in Figure~\ref{fig:Pdepo_timeseries} are not statistically significant for either filter (p-values $> 0.05$), indicating no evidence for monotonic trends in polarization over time. The polarization degree presented a variability of $\Delta P \lesssim 0.2 \%$ ($1\sigma$) in both filters, in agreement with that found using bright stars in Appendix \ref{max_SNR}. However, we note that the scatter increases to $\Delta P \approx 0.3-0.8\%$ ($1\sigma$) when no correction for ellipticity is applied.

\subsubsection{Polarization angle offset}
\label{angle_offset}

\begin{figure*}
\begin{center}
\includegraphics[width=\textwidth]{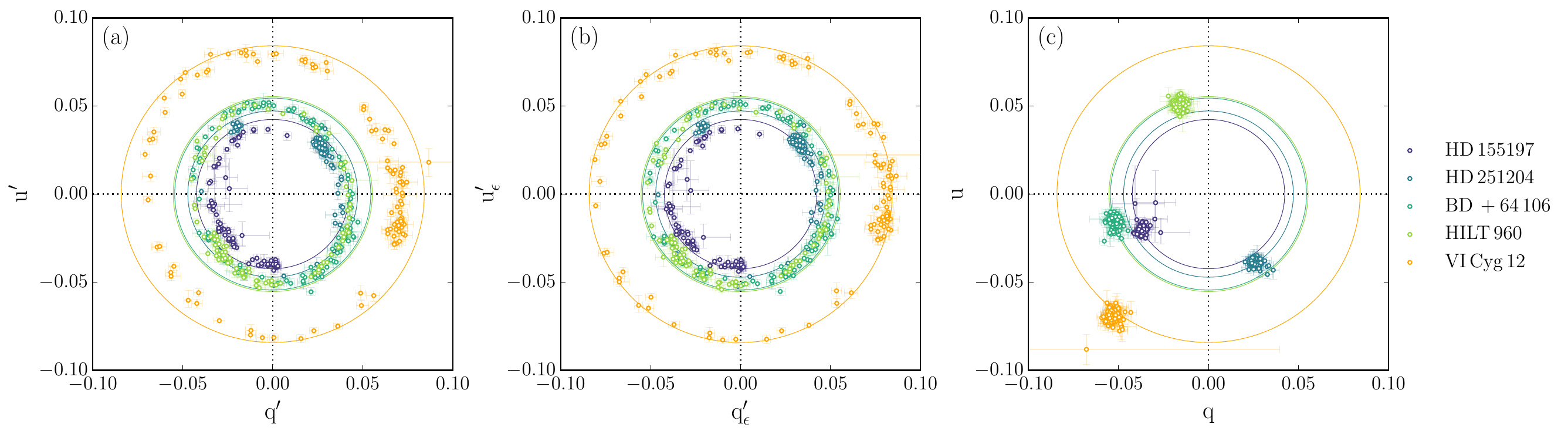}
\end{center}
\caption{Stokes parameters of the polarized standards as measured with the MOPTOP and the MOP-L filter. Solid circles correspond to the catalogue polarization degree \citep{1990AJ.....99.1243T,1992AJ....104.1563S,1999AcA....49...59W}, and dotted lines mark the $(q,u) = (0,0)$. The three figures show (a) data that have only been corrected for the instrumental offset ($q_{\rm inst}$, $u_{\rm inst}$), (b) data that have been corrected for ellipticity ($a$, $b$, $\psi$), and (c) data that have additionally been corrected by depolarization ($D_{\epsilon}$) as well as the angle rotation induced by the Liverpool Telescope ($\theta_{\rm tel}$) and MOPTOP ($\theta_{\rm inst}$).}
\label{fig:Pellipt_MOPL}
\end{figure*}

\begin{figure*}
\begin{center}
\includegraphics[width=\textwidth]{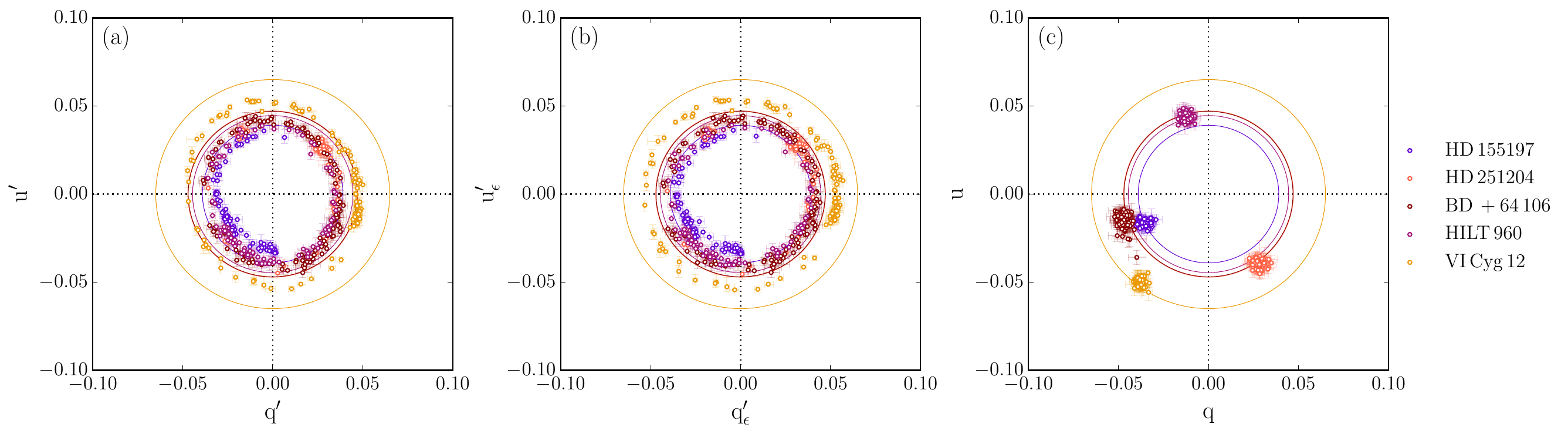}
\end{center}
\caption{The Stokes parameters of the polarized standards as measured with the MOPTOP and the MOP-I filter. See the caption of Figure \ref{fig:Pellipt_MOPL} for more details.}
\label{fig:Pellipt_MOPI}
\end{figure*}

\begin{figure*}
\begin{center}
\includegraphics[width=\columnwidth]{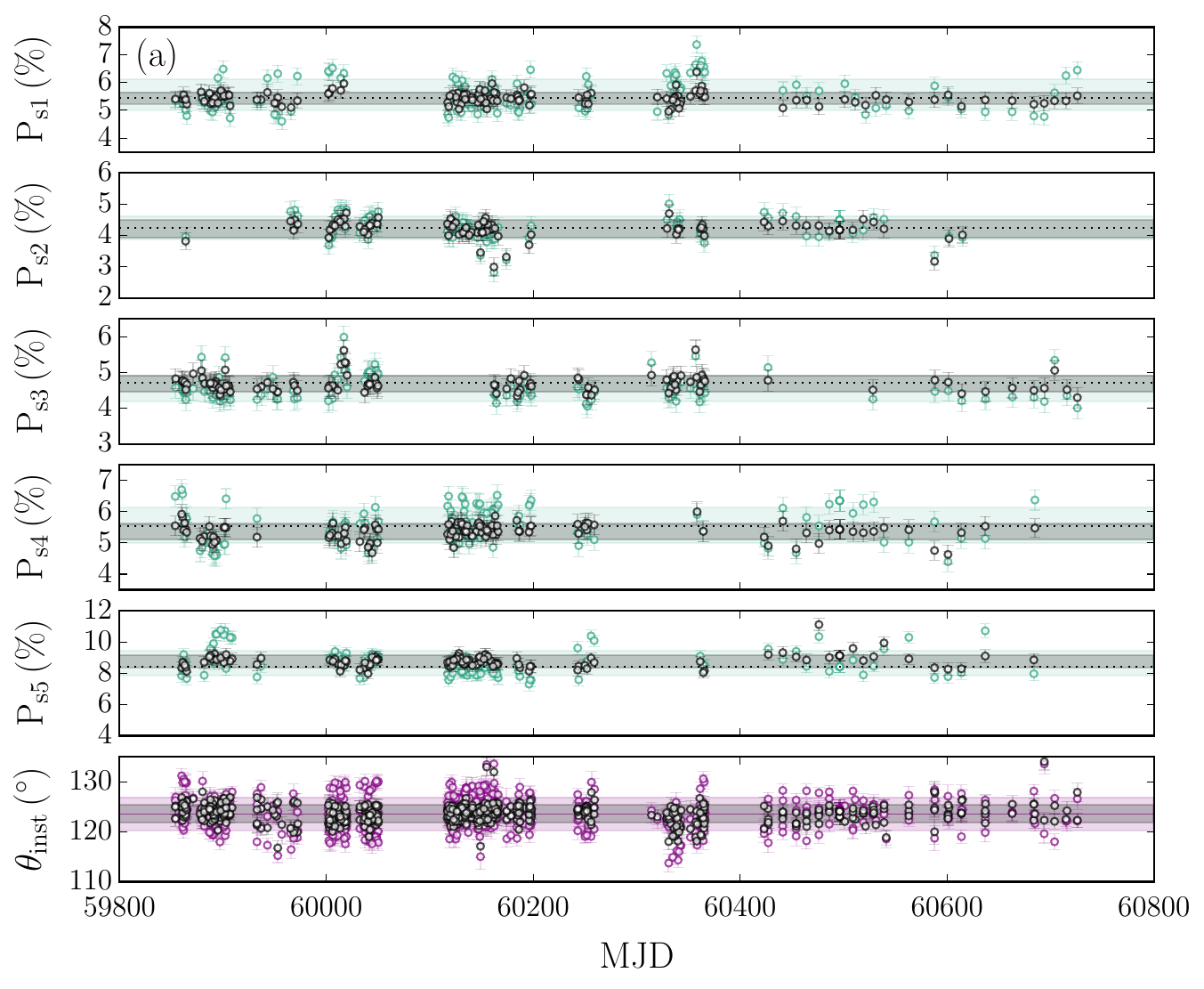}
\includegraphics[width=\columnwidth]{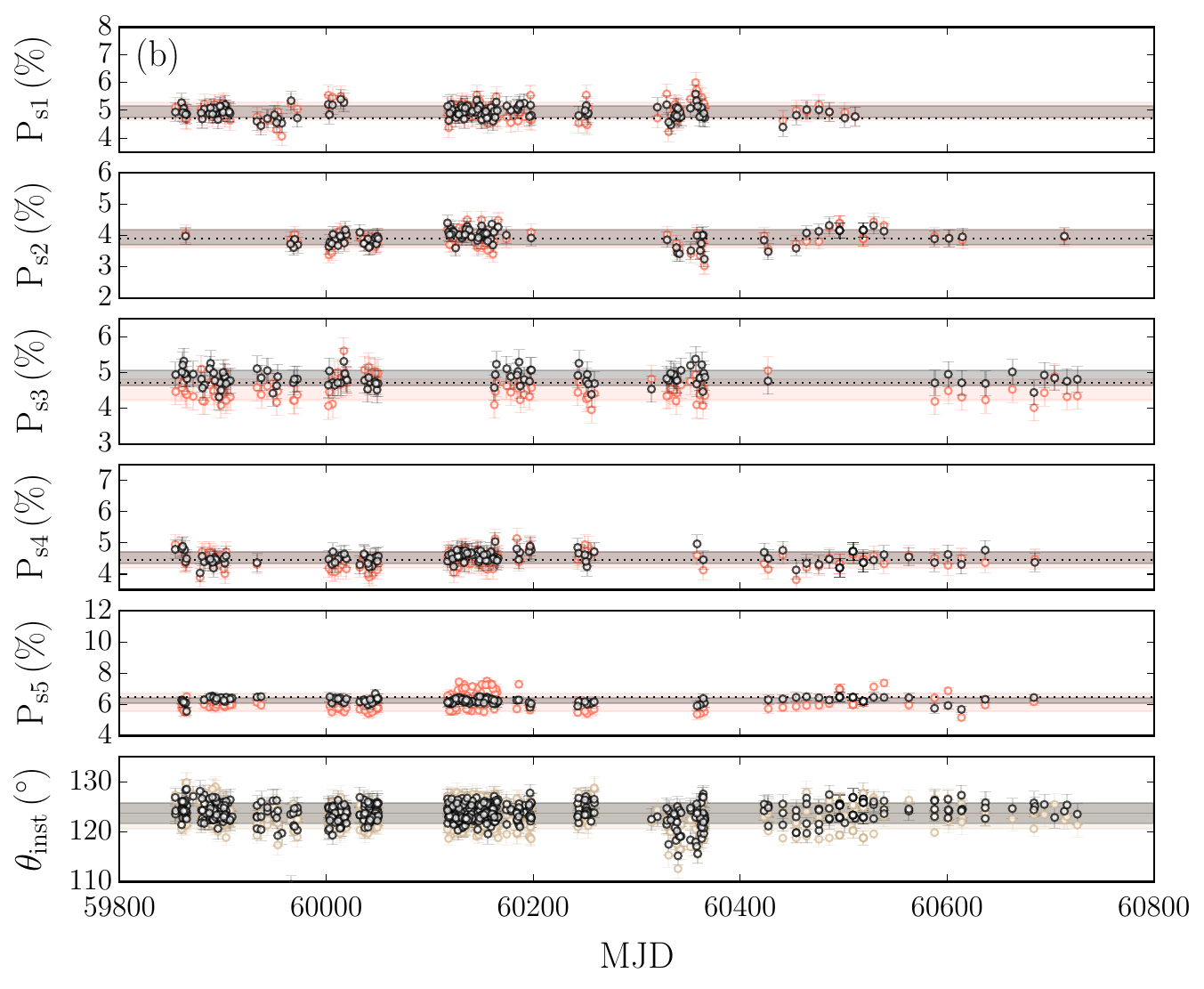}
\end{center}
\caption{The upper five panels correspond to the polarization-degree time series of the following polarized standards: BD +64 106 ($P_{s1}$), HD 155197 ($P_{s2}$), HD 251204 ($P_{s3}$), HILT 960 ($P_{s4}$), and VI Cyg 12 ($P_{s5}$). Dotted black lines indicate the catalogue polarization and the shaded regions are the standard deviation around the mean. In the bottom panels, we show the instrumental offset of the polarization angle ($\theta_{\rm inst}$); shaded regions are the standard deviation around the mean (i.e. the solid-coloured line). Note that (a) is data from the MOP-L filter and that (b) is that of the MOP-I filter. Coloured-marker data do not include ellipticity corrections, and black-marker data have all instrumental corrections implemented.}
\label{fig:Pdepo_timeseries}
\end{figure*}

Finally, we used the polarized standards to correct for the instrumental effects on the observed polarization angle ($\theta_{\rm obs}$). That is, we first corrected each observation for the rotation angle of the telescope ($\theta_{\rm tel}$), and then we determined the instrumental offset relative to the North ($\theta_{\rm inst}$) per filter. We measured an average ${\bar\theta}_{\rm inst, \, \rm L}= 124 \pm 2 \, \degr$ and ${\bar\theta}_{\rm inst, \, \rm I}= 124 \pm 2 \, \degr$ for the instrumental polarization angle; see bottom panels of Figures \ref{fig:Pdepo_timeseries}. We found significant Pearson's correlation coefficients with p-values$\,<0.05$ and $|r|=0.2$, indicating a weak but consistent monotonic decrease in polarization over time for both filters. As a result, the science observations were corrected for a local average, following the same approach as for the instrumental Stokes offset correction. The Stokes parameters were then corrected as

\[ \begin{pmatrix}
q \\
u
\end{pmatrix} 
=R(2\theta)
\begin{pmatrix}
q^{\prime \prime} \\
u^{\prime \prime}
\end{pmatrix}, \,\, {\rm where} \,\, 
R(2\theta) =
\begin{pmatrix}
\cos(2\theta) & -\sin(2\theta) \\
\sin(2\theta) & \cos(2\theta)
\end{pmatrix} \] is the rotation matrix that corrects for $\theta = \theta_{\rm tel} + \theta_{\rm inst}$; see results in Figures \ref{fig:Pellipt_MOPL}-\ref{fig:Pellipt_MOPI} (c).

\begin{table*}
\caption{MOPTOP observing log for the TDE candidates. Each entry lists the observing date, Modified Julian Date (MJD), filter used, number of polarimetric rotation cycles ($N_{\rm tot}$), airmass, seeing, fractional lunar illumination, the angular distance between the target and the Moon, and the Moon’s altitude at the time of observation. Note that the total number of frames per observation scales as $N_{\rm tot} = 32 \,N_{\rm rot}$.}       
\label{table:log}      
\centering                         
\begin{tabular}{c c c c c c c c c }
\hline               
 Observing & time & Filter & $N_{\rm rot}$ & airmass & seeing & Moon frac. & Moon dist. & Moon  alt.\\
 date &  (MJD) & & &  & ($\arcsec$) & & ($\degr$) & ($\degr$) \\
\hline
AT2019aalc \\
\hline
14/06/23 & $60109.945 \pm 0.003$ & L & 11 & 1.10 & $1.5 \pm 0.1$ & 0.1 & 156.8  & -44.4 \\
19/06/23 & $60114.987 \pm 0.003$ & L & 11 & 1.12 & $1.1 \pm 0.1$ & 0.0 & 114.7  & -18.8 \\
24/06/23 & $60120.010 \pm 0.003$ & L & 11 & 1.21 & $2.2 \pm 0.3$ & 0.4 & 60.7  & 6.3 \\
06/07/23 & $60131.932 \pm 0.003$ & L & 11 & 1.11 & $1.4 \pm 0.2$ & 0.8 & 106.9  & -13.0\\
12/07/23 & $60137.936 \pm 0.003$ & L & 11 & 1.14 & $1.2 \pm 0.1$ & 0.2 & 154.7  & -38.7 \\
18/07/23 & $60143.910 \pm 0.003$ & L & 11 & 1.12 & $1.6 \pm 0.3$ & 0.0 & 96.9  & -7.5 \\
23/07/23 & $60148.924 \pm 0.003$ & L & 11 & 1.18 & $1.3 \pm 0.2$ & 0.3 & 44.2  & 19.1 \\
03/08/23 & $60159.900 \pm 0.003$ & L & 11 & 1.20 & $1.2 \pm 0.1$ & 0.9 & 114.7  & -4.5 \\
08/08/23 & $60164.917 \pm 0.003$ & L & 11 & 1.34 & $1.2 \pm 0.1$ & 0.4 & 155.5  & -29.3 \\
20/01/24 & $60330.272 \pm 0.006$ & L & 22 & 1.32 & $2.5 \pm 0.5$ & 0.8 & 147.0  & -18.4 \\
05/02/24 & $60346.238 \pm 0.005$ & L & 19 & 1.28 & $1.7 \pm 0.2$ & 0.2 & 47.8  & 4.1 \\
06/03/24 & $60376.116 \pm 0.003$ & L & 11 & 1.50 & $1.9 \pm 0.3$ & 0.1 & 78.1  & -35.5 \\
02/04/24 & $60403.101 \pm 0.003$ & L & 11 & 1.18 & $2.1 \pm 0.3$ & 0.4 & 73.6  & -13.6 \\
11/05/24 & $60441.999 \pm 0.003$ & L & 11 & 1.17 & $1.5 \pm 0.2$ & 0.2 & 121.1  & -1.1 \\
28/05/24 & $60458.937 \pm 0.003$ & L & 11 & 1.22 & $1.7 \pm 0.2$ & 0.7 & 90.2  & -31.1 \\
28/06/24 & $60489.922 \pm 0.003$ & L & 11 & 1.09 & $1.2 \pm 0.1$ & 0.5 & 136.2  & -39.3 \\
26/07/24 & $60517.912 \pm 0.006$ & L & 22 & 1.16 & $1.3 \pm 0.2$ & 0.6 & 144.2  & -26.3 \\
27/01/25 & $60703.212 \pm 0.006$ & L & 22 & 1.62 & $1.9 \pm 0.2$ & 0.0 & 69.4  & -31.0\\
24/02/25 & $60731.223 \pm 0.006$ & L & 22 & 1.14 & $2.3 \pm 0.3$ & 0.1 & 76.6  & -11.5\\
04/02/24 & $60345.231 \pm 0.001$ & I & 5 & 1.37 & $2.0 \pm 0.2$ & 0.3 & 37.3  & 11.6 \\
04/03/24 & $60374.149 \pm 0.003$ & I & 11 & 1.31 & $2.2 \pm 0.3$ & 0.3 & 54.0  & -4.1 \\
02/04/24 & $60403.121 \pm 0.003$ & I & 11 & 1.13 & $2.2 \pm 0.3$ & 0.4 & 73.9  & -8.1 \\
11/05/24 & $60441.988 \pm 0.003$ & I & 11 & 1.20 & $1.9 \pm 0.3$ & 0.2 & 121.3  & 1.8 \\
28/05/24 & $60458.926 \pm 0.003$ & I & 11 & 1.27 & $1.7 \pm 0.2$ & 0.7 & 90.0  & -34.5 \\
28/06/24 & $60489.933 \pm 0.003$ & I & 11 & 1.10 & $1.4 \pm 0.2$ & 0.5 & 136.4  & -36.2 \\
26/07/24 & $60517.933 \pm 0.006$ & I & 22 & 1.23 & $1.2 \pm 0.1$ & 0.6 & 144.4  & -20.3 \\
\hline
AT2024bgz \\
\hline
03/03/24 & $60372.955 \pm 0.006$ & L & 22 & 1.29 & $1.7 \pm 0.2$ & 0.5 & 106.1  & -48.2 \\
13/03/24 & $60382.981 \pm 0.006$ & L & 22 & 1.19 & $1.7 \pm 0.3$ & 0.2 & 106.9  & -3.2 \\
01/04/24 & $60401.877 \pm 0.006$ & L & 22 & 1.29 & $1.5 \pm 0.2$ & 0.5 & 125.7  & -69.0 \\
26/04/24 & $60426.886 \pm 0.006$ & L & 22 & 1.20 & $1.3 \pm 0.1$ & 0.9 & 99.5  & -17.5 \\
\hline
AT2020afhd \\
\hline
05/02/24 & $60345.868 \pm 0.006$ & L & 22 & 1.22 & $1.3 \pm 0.1$ & 0.2 & 137.6  & -78.1 \\
11/02/24 & $60351.890 \pm 0.006$ & L & 22 & 1.37 & $1.3 \pm 0.1$ & 0.1 & 57.7  & -2.7 \\
26/02/24 & $60366.843 \pm 0.002$ & L & 8 & 1.36 & $1.5 \pm 0.1$ & 0.9 & 135.3  & -12.3 \\
07/08/24 & $60530.173 \pm 0.006$ & L & 22 & 1.86 & $1.7 \pm 0.2$ & 0.1 & 129.5  & -59.7 \\
27/09/24 & $60581.086 \pm 0.006$ & L & 19 & 1.38 & $1.5 \pm 0.2$ & 0.2 & 90.7  & -15.7 \\
29/10/24 & $60613.123 \pm 0.006$ & L & 22 & 1.19 & $1.6 \pm 0.2$ & 0.0 & 141.4  & -36.3 \\
20/12/24 & $60664.933 \pm 0.006$ & L & 22 & 1.17 & $1.2 \pm 0.1$ & 0.7 & 112.3  & -17.5 \\
25/01/25 & $60700.895 \pm 0.006$ & L & 22 & 1.22 & $1.4 \pm 0.1$ & 0.1 & 137.2  & -77.8  \\
04/02/24 & $60344.887 \pm 0.006$ & I & 22 & 1.27 & $1.7 \pm 0.2$ & 0.3 & 147.7  & -83.2 \\
12/02/24 & $60352.849 \pm 0.001$ & I & 4 & 1.23 & $1.7 \pm 0.2$ & 0.1 & 45.0  & 22.8 \\
05/08/24 & $60528.205 \pm 0.006$ & I & 21 & 1.56 & $1.6 \pm 0.2$ & 0.0 & 108.7  & -39.6 \\
27/09/24 & $60581.107 \pm 0.006$ & I & 21 & 1.28 & $1.6 \pm 0.2$ & 0.2 & 91.0  & -10.1 \\
28/10/24 & $60612.105 \pm 0.006$ & I & 22 & 1.17 & $1.8 \pm 0.2$ & 0.1 & 131.5  & -30.4 \\
\hline
AT2024pvu \\
\hline
14/08/24 & $60537.091 \pm 0.006$ & L & 22 & 1.05 & $1.9 \pm 0.5$ & 0.7 & 103.5  & -2.6 \\
26/09/24 & $60579.940 \pm 0.006$ & L & 22 & 1.13 & $1.2 \pm 0.2$ & 0.3 & 113.1  & -33.7 \\
22/10/24 & $60605.902 \pm 0.006$ & L & 22 & 1.05 & $2.0 \pm 0.3$ & 0.6 & 97.9  & -16.3 \\
\hline
AT2024wsd \\
\hline
28/10/24 & $60612.190 \pm 0.006$ & L & 22 & 1.46 & $2.9 \pm 0.6$ & 0.1 & 98.9  & -5.3 \\
24/11/24 & $60639.061 \pm 0.006$ & L & 18 & 1.54 & $1.9 \pm 0.2$ & 0.3 & 95.3  & -18.1 \\
\hline
\end{tabular}
\end{table*}

\begin{table*}
\caption{Galactic ISM properties of each TDE line of sight. The Galactic reddening $E(B-V)_{\rm MW}$ has been calculated for a $5 \arcmin \times 5 \arcmin $ field statistic \citep{1998ApJ...500..525S}. The $P_{\rm  serk}$ is the polarization degree estimated from the Galactic reddening and the Serkowski law \citep{1975ApJ...196..261S}.}
\label{table:prop_env}      
\centering                    
\begin{tabular}{c c c c c c c}
\hline              
TDE & Filter & $P_{\rm MW}$ & $\theta_{\rm MW}$ & $E(B-V)_{\rm MW}$ & $P_{\rm  serk}$ \\
 & & ($\%$) & ($\degr$) & (mag) &  (\%) \\
\hline
AT2019aalc & L & $0.44 \pm 0.07$ & $76 \pm 9$ & $0.0454 \pm  0.0005$ & $ \lesssim 0.59$ \\
AT2019aalc & I & $0.4 \pm 0.2$ & $77 \pm 8$ & $0.0454 \pm  0.0005$ & $\lesssim 0.53$ \\
AT2024bgz & L  & $0.3 \pm 0.1$ & $42 \pm 9$ & $0.0447 \pm  0.0006$ & $\lesssim 0.58$ \\
AT2020afhd & L & $0.4 \pm 0.2$ & $167 \pm 6$ & $0.066 \pm  0.001$ & $\lesssim 0.85$ \\
AT2020afhd & I & $0.44 \pm 0.08$ & $162 \pm 8$ & $0.066 \pm  0.001$ & $\lesssim 0.76$ \\
AT2024pvu & L & $0.50 \pm 0.03$ & $58 \pm 4$ & $0.058 \pm  0.005$ & $\lesssim 0.75$ \\
AT2024wsd & L & $0.61 \pm 0.02$ & $60.8 \pm 0.02$ & $0.060 \pm 0.001$ & $\lesssim 0.78$ \\
\hline
\end{tabular}
\end{table*}

\begin{table*}
\caption{MOPTOP polarization measurements for the TDE candidates. Polarization values are reported both before ($P_{\rm TDE + host}$, $\theta_{\rm TDE + host}$) and after subtraction of the host galaxy’s contribution ($P_{\rm TDE}$, $\theta_{\rm TDE}$). The SNR column is the average signal-to-noise ratio at which the sources were detected in each image of the final 8-frame dataset.}         
\label{table:pol_phot_resultats}
\centering     
\begin{tabular}{c c c c c c c c c c c c c c c c}
\hline               
time & Filter & SNR & $P_{\rm TDE + host}$ & $\theta_{\rm TDE + host}$ & $P_{\rm TDE}$ & $\theta_{\rm TDE}$ \\
(MJD) & & & (\%) & ($\degr$) & (\%) & ($\degr$) \\
\hline
AT2019aalc \\
\hline
$60109.945 \pm 0.003$ & L & $453 \pm 6$ & $0.00^{+0.14}_{-0.00}$ &  $8^{+30}_{-39}$ & $0.0 ^{+0.4}_{-0.0}$ &  $8^{+30}_{-39}$ \\
$60114.987 \pm 0.003$ & L & $447 \pm 3$ & $0.00^{+0.18}_{-0.00}$ &  $36^{+58}_{-18}$ & $0.0 ^{+0.4}_{-0.0}$ &  $36^{+58}_{-18}$\\
$60120.010 \pm 0.003$ & L & $390 \pm 11$ & $0.20^{+0.17}_{-0.14}$ &  $18^{+7}_{-7}$ & $0.5 ^{+0.4}_{-0.3}$ &  $18^{+7}_{-7}$ \\
$60131.932 \pm 0.003$ & L & $501 \pm 4$ & $0.29^{+0.14}_{-0.10}$ &  $6^{+10}_{-10}$ & $0.6 ^{+0.3}_{-0.2}$ &  $6^{+10}_{-10}$ \\
$60137.936 \pm 0.003$ & L & $372 \pm 3$ & $0.19^{+0.19}_{-0.11}$ &  $4^{+16}_{-15}$ & $0.4 ^{+0.4}_{-0.2}$ &  $4^{+16}_{-15}$ \\
$60143.910 \pm 0.003$ & L & $490 \pm 6$ & $0.70^{+0.13}_{-0.11}$ &  $17^{+5}_{-5}$ & $1.5 ^{+0.3}_{-0.2}$ &  $17^{+5}_{-5}$  \\
$60148.924 \pm 0.003$ & L & $505 \pm 2$ & $0.49^{+0.13}_{-0.11}$ &  $8^{+6}_{-6}$ & $1.0 ^{+0.3}_{-0.2}$ &  $8^{+6}_{-6}$\\
$60159.900 \pm 0.003$ & L & $462 \pm 2$ & $1.09^{+0.14}_{-0.13}$ &  $16^{+3}_{-3}$ & $2.4 ^{+0.3}_{-0.3}$ &  $16^{+3}_{-3}$ \\
$60164.917 \pm 0.003$ & L & $497 \pm 3$ & $0.74^{+0.14}_{-0.12}$ &  $18^{+4}_{-4}$ & $1.7 ^{+0.3}_{-0.3}$ &  $18^{+4}_{-4}$ \\
$60330.272 \pm 0.006$ & L & $530 \pm 8$ & $0.73^{+0.12}_{-0.11}$ &  $2^{+4}_{-4}$ & $2.4 ^{+0.4}_{-0.4}$ &  $2^{+4}_{-4}$ \\
$60346.238 \pm 0.005$ & L & $524 \pm 6$ & $0.47^{+0.12}_{-0.09}$ &  $178^{+7}_{-6}$ & $1.5 ^{+0.4}_{-0.3}$ &  $178^{+7}_{-6}$ \\
$60376.116 \pm 0.003$ & L & $313 \pm 5$ & $0.77^{+0.19}_{-0.14}$ &  $137^{+7}_{-7}$ & $2.3 ^{+0.6}_{-0.4}$ &  $137^{+7}_{-7}$  \\
$60403.101 \pm 0.003$ & L & $328 \pm 6$ & $1.35^{+0.19}_{-0.16}$ &  $160^{+4}_{-4}$ & $5.2 ^{+0.7}_{-0.6}$ &  $160^{+4}_{-4}$ \\
$60441.999 \pm 0.003$ & L & $315 \pm 7$ & $0.43^{+0.21}_{-0.13}$ &  $146^{+12}_{-11}$ & $1.5 ^{+0.7}_{-0.5}$ &  $146^{+12}_{-11}$ \\
$60458.937 \pm 0.003$ & L & $338 \pm 4$ & $0.47^{+0.19}_{-0.13}$ &  $138^{+11}_{-10}$ & $1.6 ^{+0.6}_{-0.4}$ &  $138^{+11}_{-10}$ \\
$60489.922 \pm 0.003$ & L & $353 \pm 4$ & $0.66^{+0.17}_{-0.13}$ &  $173^{+7}_{-7}$ & $2.9 ^{+0.8}_{-0.6}$ &  $173^{+7}_{-7}$ \\
$60517.912 \pm 0.006$ & L & $419 \pm 6$ & $0.91^{+0.15}_{-0.13}$ &  $21^{+4}_{-4}$ & $4.4 ^{+0.7}_{-0.6}$ &  $21^{+4}_{-4}$ \\
$60703.212 \pm 0.006$ & L & $438 \pm 11$ & $0.64^{+0.14}_{-0.10}$ &  $136^{+6}_{-6}$ & $5.3 ^{+1.2}_{-0.9}$ &  $136^{+6}_{-6}$\\
$60731.223 \pm 0.006$ & L & $530 \pm 9$ & $0.89^{+0.11}_{-0.09}$ &  $164^{+4}_{-4}$ & $7.5 ^{+0.9}_{-0.8}$ &  $164^{+4}_{-4}$ \\
$60345.231 \pm 0.001$ & I & $86 \pm 2$ & $1.07^{+0.91}_{-0.53}$ &  $170^{+15}_{-14}$ & $3.3 ^{+2.8}_{-1.6}$ &  $170^{+15}_{-14}$\\
$60374.149 \pm 0.003$ & I & $270 \pm 13$  & $1.18^{+0.27}_{-0.22}$ &  $157^{+5}_{-5}$ & $3.5 ^{+0.8}_{-0.7}$ &  $157^{+5}_{-5}$ \\
$60403.121 \pm 0.003$ & I & $244 \pm 17$ & $0.73^{+0.30}_{-0.22}$ &  $169^{+9}_{-9}$ & $2.8 ^{+1.1}_{-0.8}$ &  $169^{+9}_{-9}$ \\
$60441.988 \pm 0.003$ & I & $236 \pm 16$ & $1.02^{+0.29}_{-0.21}$ &  $135^{+7}_{-8}$ & $3.9 ^{+1.1}_{-0.8}$ &  $135^{+7}_{-8}$ \\
$60458.926 \pm 0.003$ & I & $251 \pm 8$ & $0.83^{+0.28}_{-0.21}$ &  $143^{+8}_{-8}$ & $3.3 ^{+1.1}_{-0.8}$ &  $143^{+8}_{-8}$ \\
$60489.933 \pm 0.003$ & I & $219 \pm 10$ & $0.97^{+0.31}_{-0.22}$ &  $0^{+8}_{-9}$ & $5.8 ^{+1.9}_{-1.3}$ &  $0^{+8}_{-9}$ \\
$60517.933 \pm 0.006$ & I & $329 \pm 19$ & $1.03^{+0.21}_{-0.17}$ &  $14^{+5}_{-5}$ & $6.4 ^{+1.3}_{-1.1}$ &  $14^{+5}_{-5}$ \\
\hline
AT2024bgz \\
\hline
$60372.955 \pm 0.006$ & L & $160 \pm 2$ & $0.56^{+0.44}_{-0.26}$ &  $129^{+14}_{-14}$ & $1.0 ^{+0.8}_{-0.5}$ &  $129^{+14}_{-14}$ \\
$60382.981 \pm 0.006$ & L & $133 \pm 5$ & $0.36^{+0.55}_{-0.20}$ &  $162^{+23}_{-18}$ & $0.7 ^{+1.1}_{-0.4}$ &  $162^{+23}_{-18}$ \\
$60401.877 \pm 0.006$ & L & $73 \pm 9$ & $2.14^{+0.92}_{-0.63}$ &  $142^{+10}_{-10}$ & $5.7 ^{+2.5}_{-1.7}$ &  $142^{+10}_{-10}$\\
$60426.886 \pm 0.006$ & L & $61 \pm 4$ & $0.19^{+1.27}_{-0.05}$ &  $156^{+25}_{-29}$ & $0.2 ^{+1.3}_{-0.1}$ &  $156^{+25}_{-29}$\\
\hline
AT2020afhd \\
\hline
$60345.868 \pm 0.006$ & L & $320 \pm 4$ & $1.34^{+0.21}_{-0.19}$ &  $2^{+4}_{-4}$ & $1.8 ^{+0.3}_{-0.3}$ &  $7 ^{+2}_{-2}$ \\
$60351.890 \pm 0.006$ & L & $230 \pm 5$ & $1.67^{+0.29}_{-0.26}$ &  $5^{+4}_{-4}$ & $2.3 ^{+0.4}_{-0.4}$ &  $9 ^{+2}_{-2} $ \\
$60366.843 \pm 0.002$ & L & $176 \pm 2$ & $1.14^{+0.39}_{-0.31}$ &  $8^{+7}_{-7}$ & $1.5 ^{+0.5}_{-0.5}$ &  $14 ^{+5}_{-5}$ \\
$60530.173 \pm 0.006$ & L & $158 \pm 9$ & $0.40^{+0.45}_{-0.18}$ &  $131^{+18}_{-20}$ & $1.8 ^{+0.5}_{-0.5}$ &  $97 ^{+17}_{-17}$ \\
$60581.086 \pm 0.006$ & L & $230 \pm 5$ & $0.56^{+0.31}_{-0.22}$ &  $100^{+10}_{-11}$ & $4.7 ^{+1.3}_{-1.3}$ &  $87 ^{+3}_{-3}$ \\
$60613.123 \pm 0.006$ & L & $316 \pm 7$ & $0.00^{+0.20}_{-0.00}$ &  $158^{+54}_{-31}$ & $3.1 ^{+0.7}_{-0.7}$ &  $75 ^{+6}_{-6}$ \\
$60664.933 \pm 0.006$ & L & $231 \pm 5$ & $0.38^{+0.31}_{-0.19}$ &  $135^{+13}_{-15}$ & $2.6 ^{+0.7}_{-0.7}$ &  $94 ^{+14}_{-14}$ \\
$60700.895 \pm 0.006$ & L & $69 \pm 2$ & $1.01^{+1.05}_{-0.59}$ &  $176^{+16}_{-15}$ & $2.6 ^{+4.1}_{-2.6}$ &  $9 ^{+18}_{-18}$ \\
$60344.887 \pm 0.006$ & I & $142 \pm 3$ & $1.56^{+0.51}_{-0.40}$ &  $2^{+8}_{-7}$ & $2.2 ^{+0.8}_{-0.8}$ &  $6 ^{+3}_{-3} $ \\
$60352.849 \pm 0.001$ & I & $79 \pm 2$ & $1.92^{+0.95}_{-0.69}$ & $2^{+10}_{-10}$ & $2.7 ^{+1.3}_{-1.3} $ &  $4 ^{+3}_{-3}$\\
$60528.205 \pm 0.006$ & I & $90 \pm 10$ & $0.10^{+0.80}_{-0.49}$ &  $118^{+12}_{-12}$ & $5.2 ^{+3.3}_{-3.3} $ &  $79 ^{+24}_{-24} $ \\
$60581.107 \pm 0.006$ & I & $153 \pm 4$ & $0.14^{+0.54}_{-0.05}$ &  $102^{+24}_{-24}$ & $7.7 ^{+3.2}_{-3.2}$ &  $79 ^{+10}_{-10} $ \\
$60612.105 \pm 0.006$ & I & $185 \pm 3$ & $0.69^{+0.39}_{-0.24}$ &  $165^{+13}_{-12}$ & $0.1 ^{+3.3}_{-0.1}$ & $-$ \\
\hline
AT2024pvu \\
\hline
$60537.091 \pm 0.006$ & L & $226 \pm 7$ & $0.69^{+0.14}_{-0.09}$ &  $137^{+10}_{-11}$ & $1.3 ^{+0.3}_{-0.3} $ &  $151 ^{+6}_{-6} $ \\
$60579.940 \pm 0.006$ & L & $310 \pm 5$ & $0.60^{+0.10}_{-0.07}$ &  $123^{+9}_{-9}$ & $1.4 ^{+0.8}_{-0.8}$ &  $147 ^{+16}_{-16} $ \\
$60605.902 \pm 0.006$ & L & $288 \pm 13$ & $0.63^{+0.22}_{-0.16}$ &  $111^{+9}_{-8}$ & $0.1  ^{+0.2}_{-0.1}$ &  $-$ \\
\hline
AT2024wsd \\
\hline
$60612.190 \pm 0.006$ & L & $51 \pm 5$ & $1.55^{+1.41}_{-0.73}$ &  $168^{+17}_{-15}$ & $2.5 ^{+2.3}_{-1.2}$ &  $168^{+17}_{-15}$\\
$60639.061 \pm 0.006$ & L & $42 \pm 1$ & $0.00^{+1.53}_{-0.00}$ &  $153^{+27}_{-58}$ & $0.0 ^{+1.5}_{-0.0}$ &  $153^{+27}_{-58}$\\
\hline
\end{tabular}
\end{table*}

\begin{table*}
\centering
\caption{MOSFiT results for the TDEs analysed in this work (and AT2020mot). Using ZTF data, each fit used at least 200 walkers and required more than $3\times 10^{4}$ iterations to achieve a potential scale reduction factor (PSRF) $ \leq 1.3$, except for AT2024wsd, which converged with $\mathrm{PSRF} = 1.35$. The model parameters show the $1\sigma$ percentile ranges derived from the marginalized posteriors; the physical parameters include the normalization of the photospheric radius ($R_{\rm ph,0}$), viscous timescale ($T_{\nu}$), mass of the SMBH ($M_{\rm BH}$), scaled impact parameter ($b$), radiative efficiency from mass fallback accretion ($\epsilon$), photospheric scaling coefficient ($l_{\rm ph}$), hydrogen column density of the host galaxy ($N_{\rm H}$), mass of the star ($M_{\star}$), and rest-frame rise time to peak bolometric luminosity in days ($t_{\rm peak}$).}
\label{tab:mosfit}
\begin{tabular}{lcccccccccc}
\hline
Name & $\log{R_{\rm ph,0}}$ & $\log{T_{\nu}}$ & $b$ & $\log{(M_{\rm BH}/M_{\odot})}$ & $\log{\epsilon}$ & $l_{\rm ph}$ & $\log{N_{\rm H}}$ & $M_{\star}/M_{\odot}$ & $t_{\rm peak}$ \\
\hline
AT2019aalc (flare 1) & $0.71^{+0.29}_{-0.26}$ & $1.89^{+0.07}_{-0.10}$ & $1.00^{+0.20}_{-0.21}$ & $6.86^{+0.21}_{-0.26}$ & $-2.41^{+0.46}_{-0.31}$ & $0.14^{+0.13}_{-0.10}$ & $19^{+2}_{-2}$ & $0.24^{+0.19}_{-0.16}$ & $113^{+5}_{-5}$ \\
AT2019aalc (flare 2) & $1.16^{+0.31}_{-0.36}$ & $1.78^{+0.13}_{-0.47}$ & $1.21^{+0.48}_{-0.58}$ & $6.22^{+0.52}_{-0.55}$ & $-1.64^{+0.50}_{-0.82}$ & $0.45^{+0.23}_{-0.21}$ & $19^{+2}_{-2}$ & $0.08^{+0.41}_{-0.05}$ & $99^{+8}_{-8}$\\
AT2020afhd (flare 2) & $0.52^{+0.28}_{-0.28}$ & $-0.19^{+1.10}_{-2.05}$ & $1.02^{+0.19}_{-0.14}$ & $6.66^{+0.20}_{-0.43}$ & $-2.55^{+0.52}_{-0.51}$ & $0.27^{+0.13}_{-0.12}$ & $19^{+2}_{-2}$ & $0.15^{+0.32}_{-0.11}$ & $49^{+2}_{-2}$ \\
AT2024pvu & $2.31^{+0.49}_{-0.57}$ & $0.17^{+0.71}_{-2.08}$ & $0.86^{+0.12}_{-0.08}$ & $7.20^{+0.12}_{-0.14}$ & $-2.91^{+0.37}_{-0.33}$ & $1.93^{+0.05}_{-0.10}$ & $20^{+1}_{-3}$ & $1.88^{+2.12}_{-0.79}$ & $81^{+7}_{-2}$\\
AT2024wsd & $0.54^{+0.19}_{-0.16}$ & $-0.38^{+1.17}_{-1.74}$ & $0.91^{+0.34}_{-0.26}$ & $6.39^{+0.13}_{-0.33}$ & $-2.57^{+0.50}_{-0.47}$ & $0.65^{+0.14}_{-0.12}$ & $19^{+1}_{-2}$ & $0.32^{+0.33}_{-0.17}$ & $32^{+2}_{-3}$\\
AT2024bgz & $0.82^{+0.08}_{-0.09}$ & $-0.28^{+0.79}_{-1.31}$ & $1.03^{+0.05}_{-0.07}$ & $6.55^{+0.10}_{-0.18}$ & $-2.52^{+0.43}_{-0.27}$ & $0.43^{+0.06}_{-0.05}$ & $18^{+2}_{-2}$ & $0.19^{+0.15}_{-0.12}$ & $39^{+3}_{-3}$ \\
AT2020mot & $0.60^{+0.16}_{-0.15}$ & $-0.29^{+1.27}_{-1.66}$ & $0.76^{+0.10}_{-0.10}$ & $7.12^{+0.08}_{-0.05}$ & $-3.21^{+0.09}_{-0.10}$ & $0.44^{+0.09}_{-0.08}$ & $18^{+2}_{-1}$ & $1.49^{+0.35}_{-0.20}$ & $89^{+3}_{-5}$\\
\hline
\end{tabular}
\end{table*}

\end{appendix}

\end{document}